\documentclass[10pt,journal]{IEEEtran}
\usepackage{amsmath,amssymb,amsfonts}
\usepackage{algorithmic}
\usepackage{array}
\usepackage{bm}
\usepackage[utf8]{inputenc}
\usepackage{graphicx}
\usepackage[ruled,linesnumbered,lined]{algorithm2e}
\usepackage{makecell}
\usepackage{diagbox}
\usepackage{subfigure}
\usepackage{cite}
\usepackage{booktabs}
 \usepackage{multirow} 
\usepackage[hidelinks]{hyperref}

\begin{document}
\title{Simultaneously Transmitting and Reflecting Reconfigurable Intelligent Surfaces Empowered  Cooperative Rate Splitting with User Relaying} 
\author{ Kangchun Zhao, Yijie Mao, \textit{Member, IEEE}, Yuanming Shi \textit{Senior Member, IEEE}
\vspace{-8mm}
\thanks{A preliminary version of this paper was presented at the IEEE 98th Vehicular Technology Conference, VTC2023-Fall, 2023 \cite{zhao2023star}. }
\thanks{This work has been supported in part by the National Nature Science Foundation of China under Grant 62201347; and in part by Shanghai Sailing Program under Grant 22YF1428400.}
\thanks{K. Zhao, Y. Mao, and Y. Shi are with the School of Information Science and Technology, ShanghaiTech University, Shanghai 201210, China (e-mail: zhaokch12022@shanghaitech.edu.cn, maoyj@shanghaitech.edu.cn, shiym@shanghaitech.edu.cn).}
\thanks{}}

\maketitle
\begin{abstract}
In this work, we unveil the advantages of synergizing cooperative rate splitting (CRS) with user relaying and  simultaneously transmitting and reflecting reconfigurable intelligent surface (STAR RIS).
Specifically, we propose a novel STAR RIS-assisted CRS transmission framework, featuring six unique transmission modes that leverage various combination of the relaying protocols (including full duplex-FD and half duplex-HD) and the STAR RIS configuration protocols (including energy splitting-ES, mode switching-MS, and time splitting-TS).
With the objective of  maximizing the minimum user rate, we then propose a unified successive convex approximation (SCA)-based alternative optimization (AO) algorithm to jointly optimize the transmit active  beamforming, common rate allocation, STAR RIS passive beamforming, as well as time allocation (for HD or TS protocols) subject to the transmit power constraint at the base station (BS) and the  law of energy conservation at the  STAR RIS.
To alleviate the computational burden, we further propose a low-complexity algorithm that incorporates a closed-form passive beamforming design.
Numerical results show that our proposed framework significantly enhances user fairness compared with conventional CRS schemes without STAR RIS or other STAR RIS empowered multiple access schemes. 
Moreover, the proposed low-complexity algorithm dramatically reduces the computational complexity while achieving very close performance to the AO method. 
\end{abstract}


\section{Introduction}
In sixth-generation (6G) and beyond mobile communications, an increasing number of devices are connecting to the wireless network, causing the following  problems: 1) Multi-device connections  introduce significant multi-user interference,  severely  degrading system performance. 2)  Ensuring a favorable communication environment for each user becomes highly challenging  due to the vast number of  devices accessing to the transmission network.
To deal these  problems,  a novel non-orthogonal  interference management strategy in the physical (PHY) layer named rate-splitting multiple access (RSMA) comes into our sight. It emerges as a promising technique for enhancing the spectral efficiency and user fairness \cite{clerckx2023primer}. 
RSMA follows the design principle where user messages are divided into common and private parts at the transmitter.  The common parts are encoded into common streams and decoded by multiple users. Meanwhile, the private parts are independently encoded into private streams and  decoded only by the corresponding users, following the decoding of the common streams and their subsequent removal through successive interference cancellation (SIC).
This empowers RSMA to partially treat interference as noise and partially decode interference. It encompasses various
 multiple access (MA) schemes like space division MA (SDMA), which fully treats interference as noise, and non-orthogonal MA (NOMA), which fully decodes interference, treating them as special cases  \cite{mao2017rate,mao2019rate,clerckx2019rate}.
\par
However, as the common streams of RSMA are required to be decoded by multiple  users,  the achievable  rate will be limited by the user with the  weakest channel strength.
To overcome this limitation, a novel RSMA scheme named cooperative rate splitting (CRS) is proposed in \cite{jian2019crs,Mao2020}. This scheme empowers users with stronger channel conditions, acting as relaying users, to transmit the decoded common stream to users with weaker channel conditions. 
CRS thereby boosts  the received signal strengths at the weaker users and improves the achievable common rate. 
It not only enhances spectral efficiency but also extends radio coverage, promotes user fairness, and reduces energy consumption \cite{Mao2020,khisa2022full}. There are two transmission phases in CRS, one is the direct transmission phase  that the base station (BS) transmits signals to all users, the other is the cooperative transmission phase that the relaying users transmit the decoded common streams to the destination users. According to whether these two phases are executed at the same time, there are two protocols in CRS: \textit{half duplex (HD)} \cite{Mao2020} and  \textit{full duplex (FD)}\cite{li2021full}. In the HD protocol, the direct  and cooperative transmission phases are executed in different time slots, while in the FD protocol, these two transmission phases are executed at the same time. Both protocols demonstrate significant potential in realizing the aforementioned advantages of CRS.

In parallel, reconfigurable intelligent surface (RIS), gaining increasing  attention recently,  has been viewed as a promising technique in future wireless networks \cite{pan2021differential,basar2019wireless,li2023reconfigurable}.
A RIS is a two-dimensional (2D) meta-surface  containing numerous passive and low-cost reflecting elements which enable to tune the phase shift of the incident signal.
By employing the RIS, the line-of-sight (LoS) channel between the BS and users can be reconfigured,  effectively improving the spectral efficiency, energy efficiency and extending the communication coverage.
Nevertheless, conventional RIS can solely reflect signals, limiting its coverage to a $180^{\circ}$ area.
  This has motivated the emergence of a novel RIS that enables to reflect and transmit the incident signal at the same time, which is called  simultaneously transmitting and reflecting RIS (STAR RIS) \cite{liu2021star} or intelligent omni-surface
(IOS) \cite{zhang2022intelligent}.
According to different operating methods, STAR RIS has three protocols: \textit{energy splitting (ES)}, \textit{mode switching (MS)}, and \textit{time switching (TS)}.
In the ES protocol, all elements can reflect and transmit the incident signal at the same time. 
In the MS protocol, each element either reflects or transmits the incident signal.
In the TS protocol, the transmission time is split into two time slots. One is the reflection time slot, all elements reflect the incident  signal, the other is the transmission time slot, all elements transmit the incident signal.
By  adjusting the amplitude and phase shift of the reflected and transmitted signals, all three protocols of STAR RIS offer  full spatial coverage and additional degrees of freedom (DoF) for the system. 

\par 
In view of the advantages of the STAR RIS, many existing works have explored its integration with various MA schemes 
\cite{wu2021coverage,zhang2022secrecy,zuo2022joint,hashempour2022secure,katwe2023improved,dhok2022rate}.
In \cite{wu2021coverage}, the authors investigated the coverage range of a two-user STAR RIS-aided orthogonal MA (OMA) or NOMA system.
In \cite{zhang2022secrecy}, the beamforming optimization was studied in a uplink STAR RIS-aided NOMA system for  outage probability minimization and secrecy capacity maximization.
In \cite{zuo2022joint}, the sum rate maximization was investigated in a downlink STAR RIS-aided NOMA system, wherein the decoding order and transmit beamforming are jointly optimized.
Besides, the integration of STAR RIS and RSMA  has recently been investigated in some existing works \cite{hashempour2022secure,katwe2023improved,dhok2022rate}.
In \cite{hashempour2022secure}, the primary focus was on maximizing the sum secrecy rate in a STAR RIS-aided simultaneous wireless information and power transfer (SWIPT) system using RSMA.
In \cite{katwe2023improved}, the sum-rate maximization problem for a STAR RIS-aided uplink RSMA system is studied. 
In \cite{dhok2022rate}, efforts were directed towards minimizing the outage probability while considering the spatial correlation among the STAR RIS channels.
However, so far, the integration of STAR RIS and CRS has not been investigated yet. And the performance of max-min fairness for STAR RIS-aided RSMA remains unexplored.

Considering the aforementioned advantages of both CRS and STAR RIS, along with the identified  research gaps in current literature, there is a compelling incentive to integrate these two techniques.
On one hand, the CRS transmission scheme provides robust and flexible  interference management to enhance the performance of STAR RIS. On the other hand, STAR RIS reconfigures and improves wireless channels, thereby further enhancing the performance of CRS. Their integration can lead to a mutually beneficial solution.
In this work, we delve into the STAR RIS-aided CRS system, and the main contributions of this paper are summarized as follows:
\begin{itemize}
    \item We propose a novel downlink STAR RIS-aided CRS transmission framework, empowering a STAR RIS to assist both the direct and  cooperative transmission phases of CRS. Within this framework, we investigate six different transmission modes including various combinations of CRS relaying protocols (HD and FD) and STAR RIS operating protocols (ES, MS, and TS).
    \item 
    We formulate a new resource allocation problem with the objective of maximizing the minimum user rate.
    To solve this problem,  the  STAR RIS passive  beamforming, the BS active  beamforming, common rate allocation, and  time slot allocation (in HD or TS protocols)  are jointly optimized under the transmit power constraint at the BS and the energy conservation constraints at the STAR RIS.
    Due to the non-convexity of the formulated problem, we propose an alternative optimization (AO) algorithm to solve the problem. This approach involves decomposing the original problem into two subproblems:  the STAR RIS passive beamforming optimization  and  transmit active beamforming optimization. Each subproblem is then solved using a successive convex approximation (SCA)-based method. Through iterative solving of the two subproblems, we attain a near-optimal solution until convergence.
    \item 
    We further  propose a low-complexity algorithm to solve the formulated problem. For the passive beamforming design, we derive a closed-form solution for the STAR RIS passive beamforming based on the gradient decent approach. To ensure the derived solution meets the STAR RIS constraints, we further use the symmetric unitary projection  based on singular value decomposition (SVD) to project the solution into the feasible set of the constraints. For transmit active beamforming, we use the zero-forcing (ZF) approach to fix the beamforming direction. Subsequently, we simply optimize the power allocation using SCA.
    \item
    We evaluate the performance of the proposed STAR RIS-aied CRS framework and  show the effectiveness of our proposed  algorithms by numerical results. 
    Our analysis reveals that the STAR RIS-aided CRS scheme outperforms other STAR RIS-aided MA schemes. We also offer insights into the preferred regions for the  six proposed transmission protocols. Moreover, the results  demonstrate that our proposed low-complexity algorithm achieves comparable performance while significantly reducing CPU time compared to the  AO algorithm.
\end{itemize} 
\par\textit{Organization}: The subsequent sections of the paper are organized as follows. Section \ref{system} delineates the system model and formulates the max-min fairness problem. 
Section \ref{optimization framework} details the AO optimization framework for the problem and the proposed low-complexity algorithm.
Section \ref{numerical results} presents numerical results.
And Section \ref{conclusion} concludes the paper.

\section{SYSTEM MODEL and PROBLEM FORMULATION}
\label{system}

\begin{figure}[tb]
\centerline{\includegraphics[width=7.9cm]{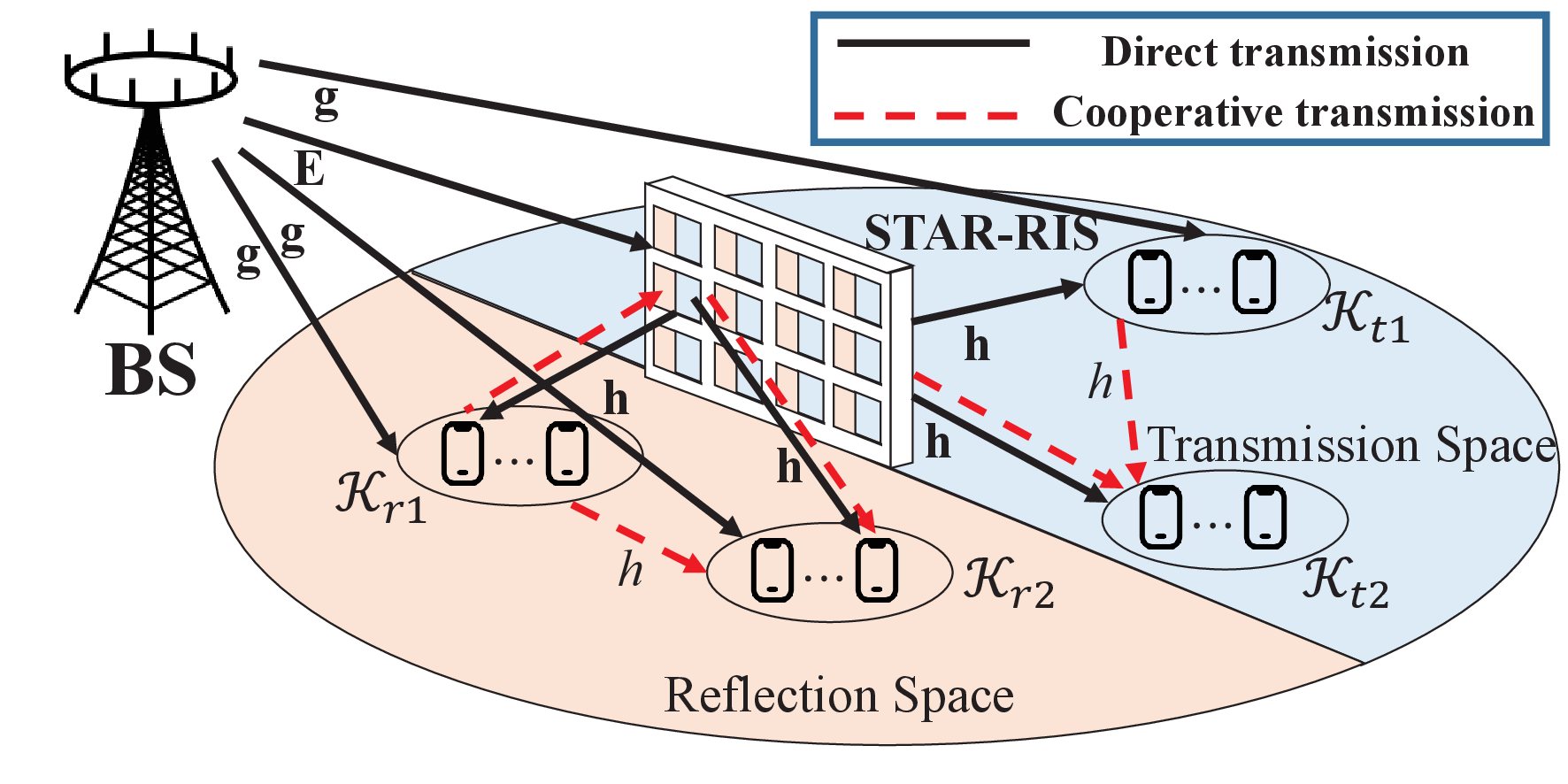}}
\vspace{-0.3cm}
\caption{The transmission architecture of the proposed STAR RIS-assisted CRS.}
\label{fig1}
\vspace{-0.3cm}
\end{figure}

As illustrated in Fig. \ref{fig1}, we consider a STAR RIS-assisted multi-user multiple-input single-output (MISO) downlink transmission network, with CRS supported during the transmission. There is  a BS equipped with $L$  transmit antennas,  a STAR RIS containing $N$ elements indexed by the set $\mathcal{N} = \{1, \cdots ,N\}$, and $K$   users indexed by the set $\mathcal{K}=\{1,2,\cdots,K\}$. Each user is equipped with a single transmit antenna and a receive antenna.
Given the ability of the STAR RIS  to  transmit and reflect the incident signal simultaneously, we divide the transmission space into two distinct subspaces, namely, the reflection space and the transmission space. 
Users in the reflection space receive signals reflected by the STAR RIS and are indexed by the set  $\mathcal{K}_r$.
Conversely, users in the transmission space receive
signals transmitted by the STAR RIS and  are indexed by the set $\mathcal{K}_t$. 
The two user sets satisfy   $\mathcal{K}_r\cup \mathcal{K}_t=\mathcal{K}$ and $\mathcal{K}_r\cap \mathcal{K}_t=\emptyset$.
Both subspaces allow user relaying, which further divides the users in each subspace into two user groups: the 
 relaying user group $\mathcal{K}_{i1}$ and the destination user group $\mathcal{K}_{i2}, i\in\{r,t\}$.
The user groups  satisfy $\mathcal{K}_{r1}\cap \mathcal{K}_{r2}=\emptyset, \mathcal{K}_{t1}\cap \mathcal{K}_{t2}=\emptyset$,  $\mathcal{K}_{r1}\cup \mathcal{K}_{r2}=\mathcal{K}_r$, and $\mathcal{K}_{t1}\cup \mathcal{K}_{t2}=\mathcal{K}_t$.
For simplicity, we denote the relaying user set  and the destination user set in the full space as  $\mathcal{K}_{1}=\mathcal{K}_{t1}\cup\mathcal{K}_{r1}$ and $\mathcal{K}_{2}=\mathcal{K}_{t2}\cup\mathcal{K}_{r2}$,
respectively.
Without loss of generality, we assume users in  $\mathcal{K}_{1}$  have  better channel conditions than users in $\mathcal{K}_{2}$.
\par
In this work, we consider a block-fading channel model with $T$ transmission blocks indexed by the set $\mathcal{T}=\{1,\cdots,T\}$. 
The channels between the BS and STAR RIS,  the BS and user-$k$, STAR RIS and user-$k$  are respectively denoted by $\mathbf{E}(t)\,\in\,\mathbb{C}^{N\times L},\,\mathbf{g}_k(t)\,\in\,\mathbb{C}^{L\times 1}$, $\mathbf{h}_{k}(t)\,\in\,\mathbb{C}^{N\times1}$, where $t$ denotes the $t$-th block, $t\in \mathcal{T}$.
Besides, the channel between  user-$i$ to user-$j$ is denoted by $h_{i,j}$.
As the pioneer study on STAR RIS-empowered CRS, we employ a simplified CSI model by assuming perfect knowledge of CSI at the BS for all communication links, i.e., perfect CSI at the transmitter (CSIT). The imperfect CSIT setting can be easily extended based on the existing works on RSMA with imperfect CSIT.  It's important to highlight that our primary emphasis is on exploring the diverse performance achieved by various transmission modes of STAR RIS-empowered CRS.


\subsection{Transmission Modes of STAR RIS-empowered CRS}
\par
CRS involves two transmission phases: the direct transmission phase and the cooperative transmission phase. During the direct transmission phase, the BS sends signals to all users. In the cooperative transmission phase, users in $\mathcal{K}_1$ act as relaying users to transmit the decoded common stream to users in $\mathcal{K}_2$, utilizing either the HD or FD protocols. The FD protocol allows simultaneous transmission and reception for relaying users. However, this approach introduces self-interference at the relaying users. In the \textit{HD protocol}, each time block is divided into two consecutive parts respectively for the two transmission phases. Let $\lambda$ $ (0 < \lambda \leq 1)$ denote the fraction of time allocated to the direct transmission phase, and $(1-\lambda)$ is the fraction allocated to the cooperative transmission phase. 

 In this paper, both transmission phases of CRS are assisted by the STAR RIS. According to the principle of single connected STAR RIS in \cite{liu2021star}, each element of STAR RIS can operate in the reflection mode and/or the transmission  mode according to different operating protocols \cite{liu2021star}. 
The reflection and transmission matrices are all  diagonal matrices, denoted as $\bm{\Theta}_i=\text{diag}\left(\sqrt{\beta_{1,i}}e^{j\theta_{1,i}},\cdots,\sqrt{\beta_{N,i}}e^{j\theta_{N,i}}\right)$, $i\in\{r,t\}$.
In the \textit{ES protocol},  each element of STAR RIS simultaneously reflects and transmits the incident signal. Hence, $\beta_{n,r}$ and $\beta_{n,t}$ should satisfy $\beta_{n,r}+\beta_{n,t}=1$ and $\beta_{n,r}, \beta_{n,t}\in[0,1]$. 
The \textit{MS protocol} is a special case of ES where each element operates  in either reflection or transmission mode, resulting in binary values for the amplitude coefficients, i.e., $\beta_{n,r}, \beta_{n,t}\in{0,1}$.
In the \textit{TS protocol}, each element of STAR RIS switches between the transmission and reflection modes in each transmission block. Define $0 \leq \lambda_r \leq 1$  as the percentage of time allocated to the reflection mode in each transmission block, $(1-\lambda_r)$  as the percentage of time allocated to the transmission mode. The TS protocol focuses on designing $\lambda_r$ and the phase shifts, where $\beta_{n,r}=1, \beta_{n,t}=0$ during the $\lambda_r$ percentage of time and  $\beta_{n,r}=0, \beta_{n,t}=1$ for the remaining  $(1-\lambda_r)$. 

\begin{figure}[!tb]
\centerline{\includegraphics[width=0.5\textwidth]{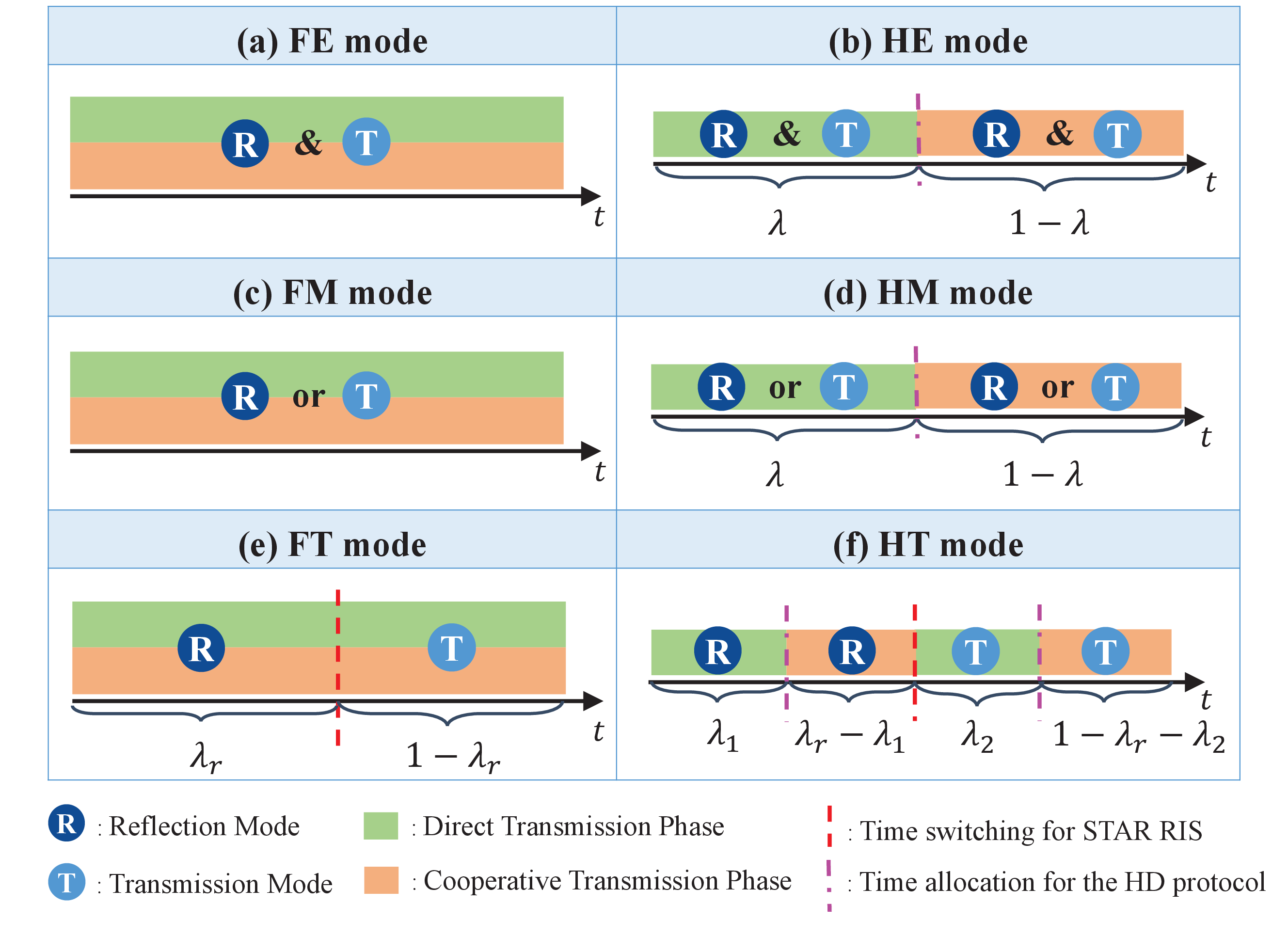}}
\vspace{-0.3cm}
\caption{Six transmission modes of the proposed STAR RIS-assisted CRS.}
\label{fig:Six modes}
\vspace{-0.3cm}
\end{figure}
\par
In the proposed STAR RIS-assisted CRS transmission framework,  we pair the aforementioned relaying protocols (HD and FD) with the three STAR RIS operating protocols (ES, MS, and TS), yielding six distinct transmission modes for STAR RIS-assisted CRS. These modes are summarized in Fig. \ref{fig:Six modes} and explained as follows: 
\begin{itemize}
\item \textit{FD-ES (FE) mode}: This refers to the use of the HD  protocol for CRS and the ES  protocol for the STAR RIS.  As illustrated in Fig. \ref{fig:Six modes}(a), HE enables each element of the STAR RIS to operate in  both transmission and reflection modes simultaneously in the non-orthogonal direct and cooperative transmission phases. 

\item \textit{HD-ES (HE) mode}: This refers to the use of the HD  protocol for CRS and the ES  protocol for the STAR RIS. 
As illustrated in Fig. \ref{fig:Six modes}(b), HE enables each element of the STAR RIS to operate in  both transmission and reflection modes simultaneously in the orthogonal direct and cooperative transmission phases. 

\item \textit{FD-MS (FM) mode}: This refers to the use of the FD  protocol for CRS  and the MS protocol for the STAR RIS.   As illustrated in Fig. \ref{fig:Six modes}(c), FM enables each element of the STAR RIS to operate in  either transmission or reflection mode in the non-orthogonal direct and cooperative transmission phases. 

\item \textit{HD-MS (HM) mode}: This refers to the use of the HD  protocol for CRS  and the MS protocol for the STAR RIS.   As illustrated in Fig. \ref{fig:Six modes}(d), HM enables each element of the STAR RIS to operate in  either transmission or reflection mode in the orthogonal direct and cooperative transmission phases. 

\item \textit{FD-TS (FT) mode}:  This refers to the use of the FD  protocol for CRS  and the TS protocol for the STAR RIS.  As illustrated in Fig. \ref{fig:Six modes}(e), each transmission block is split into two time slots based on $\lambda_r$ ($0 \leq \lambda_r \leq 1$). FT facilitates all elements of the STAR RIS to function in the reflection mode during the initial portion $\lambda_r$ of the time, while they operate in the transmission mode during the subsequent portion $1-\lambda_r$ of the time, across the non-orthogonal direct and cooperative transmission phases.

\item \textit{HD-TS (HT) mode}: This refers to the use of the HD  protocol for CRS  and the TS protocol for the STAR RIS.  As illustrated in Fig. \ref{fig:Six modes}(f), each transmission block is split into two time slots based on $\lambda_r$:  one for the reflection mode and the other for the transmission mode of STAR RIS. Then, each time slot is further split into two orthogonal parts for the direct and cooperative transmission phases of CRS, resulting in a total of four orthogonal transmission slots.
Here, we introduce $\lambda_1$ ($0 < \lambda_1 \leq \lambda_r $) and $\lambda_2$ ($0 < \lambda_2 \leq 1-\lambda_r $) to  represent the time allocation between the two transmission phases in the reflection and transmission modes, respectively.

\end{itemize}
Hereinafter, we use $j$ to represent the selected transmission mode, where $j\in \{\mathrm{FE},\mathrm{FM},\mathrm{FT},\mathrm{HE},\mathrm{HM},\mathrm{HT}\}$.

\subsection{Transmit Signal Model}
\label{signal}
For all six transmission modes  described in the previous subsection, the primary  differences among them lie in the received signals and the approaches employed  to process the received signals. 
The transmit signals in both the direct and cooperative transmission phases remain consistent, as  specified in this subsection. 
\subsubsection{Direct transmission phase}
We denote  $W_k$ as the message intended to user-$k$.
At the BS, according to the principle of 1-layer RSMA \cite{RSintro16bruno,mao2022rate}, $W_{k}$  is split into a common sub-message $W_{c,k}$ and a private sub-message $W_{p,k}$.
All  common sub-messages $W_{c,1},\cdots,W_{c,K}$ are collected and then encoded into a common stream $s_0$, which is  required to be decoded by all users.
The individual  private sub-message $W_{p,k}$ is independently encoded into a private stream $s_k$, which is required to be decoded by user-$k$ only. 
Denote  $\mathbf{s}=[s_0,s_1,\ldots,s_K]^T$ and $\mathbf{P}=[\mathbf{p}_{0},\mathbf{p}_{1},\ldots,\mathbf{p}_{K} ]\in \mathbb{C}^{L\times (K+1)}$ respectively as the stream vector and the transmit beamforming matrix, where $\mathbf{p}_{k}\in\mathbb{C}^{L\times 1}$.
The transmit signal at the BS in the direct transmission phase is given by 
\begin{equation}
\small
\mathbf{x}^{[1]}=\mathbf{P}\mathbf{s}=\sum_{k=0}^{K}\mathbf{p}_{k}s_k,
\end{equation}
where the superscript $[1]$ denotes the  direct transmission phase. Here, assuming $\mathbb{E}\{\mathbf{s}\mathbf{s}^{H}\}=\mathbf{I}$, the transmit
power of the BS is limited by  $\textrm{tr}(\mathbf{P}\mathbf{P}^H)\leq P_{t}$, where $P_{t}$ is the maximum transmit power at the BS.
\subsubsection{Cooperative transmission phase}
Users in  $\mathcal{K}_1$  employ the decode-and-forward (DF) protocol to decode, re-encode, and subsequently transmit the  decoded  $s_0$ to users in  $\mathcal{K}_2$.
The transmit signal at the relaying user-$k$ in the cooperative transmission phase is 
\begin{equation}
x^{[2]}_k=\sqrt{P_k}s_0,
\end{equation}  
where the superscript $[2]$ denotes  the cooperative transmission phase and $P_k$ is the relaying transmit power at user-$k$.
\par

\subsection{Received Signal Model for FD-based Transmission Modes}
\label{CRS FD}

In the FD-based transmission modes (FE, FM, and FT),  both the direct and cooperative transmission phases share the same radio resources.
Self-interference is inevitably among the relaying users in $\mathcal{K}_1$ as  they  simultaneously receive signals from the BS and transmit the common stream $s_0$ to  users in $\mathcal{K}_2$, \cite{Riihonen2011,Sabharwal2014}.
 We denote the self-interference channel as $I_{k}(t)$, which follows the same distribution $I_k(t)\sim  \mathcal{CN}(0, \Omega_I^{2})$ for all $ k\in \mathcal{K}_1$.
The signal received by users in $\mathcal{K}_2$ from $\mathcal{K}_1$, denoted as $\hat{s}_0(t)$, is a delayed version of $s_0$, where $\hat{s}_0(t)=s_0(t-t_d)$. Here, $t_d$ represents the processing time for users in $\mathcal{K}_1$ to decode and forward $s_0$. To ensure that the relaying users transmit and receive $\hat{s}_0(t)$ within the same block, we assume that $t_d$ is much smaller than one block period \cite{Sabharwal2014,6898012}.
 In this work, we focus on optimizing  $\mathbf{P}$, $\bm{\Theta}_r$ and $\bm{\Theta}_t$  in each transmission block. Hence, we omit $t$ in the following.
 Next, we will specify the received signals and achievable rates for FE, FM, and FT, respectively.
\subsubsection{FE/FM}

When $j=\mathrm{FE}$ or $j=\mathrm{FM}$, the signal received at each relaying user-$k$ in  $\mathcal{K}_1$ is given by
\begin{equation}
\small
\begin{split}
\label{FD-receive signal}  y^{j}_{k}=\underset{\tilde{\mathbf{g}}^H_{k,i}}{\left(\underbrace{\mathbf{g}_k^H+\mathbf{h}_{k}^H\mathbf{\Theta}_i\mathbf{E}}\right)}\mathbf{x}^{[1]}+I_{k}\sqrt{P_{k}}\hat{s}_0+n_k,i\in\{r,t\},
  \end{split}
  \end{equation}
where  $n_k$ is the additive white Gaussian noise (AWGN) which follows $n_k\sim  \mathcal{CN}(0, \sigma^{2})$ and  $I_k\sqrt{P}_k\hat{s}_0$ is the self-interference suffered at user-$k$. 
Each user-$k$ in  $\mathcal{K}_1$  sequentially decodes the common stream $s_0$ and the private stream $s_k$ with the assistance of SIC.
Let $\tilde{\mathbf{g}}_{k,i}$ denote the effective channel between the BS and user-$k$, the SINRs of decoding $s_0$ and $s_k$ at user-$k$  in  $\mathcal{K}_1$ are respectively given as
\begin{align}    
&\gamma^{\mathrm{FD}}_{c,k}=\gamma^{\mathrm{FD},[1]}_{c,k}=\frac{\left|\mathbf{\tilde{g}}_{k,i}^H\mathbf{p}_{0}\right|^2}{\sum_{m\in\mathcal{K}}\left|\mathbf{\tilde{g}}_{k,i}^H\mathbf{p}_{m}\right|^2+\left|I_{k}\right|^2P_{k}+\sigma^2},\label{Gamma_c1}\\
&\gamma^{\mathrm{FD}}_{k}=\frac{\left|\mathbf{\tilde{g}}_{k,i}^H\mathbf{p}_{k}\right|^2}{\sum_{m\in\mathcal{K},m\neq k}\left|\mathbf{\tilde{g}}_{k,i}^H\mathbf{p}_{m}\right|^2+\left|I_{k}\right|^2P_{k}+\sigma^2}.\label{Gamma_p1}
\end{align}
\par Each user-$k$ in  $\mathcal{K}_2$ receives signals from both the BS and the relaying users in $\mathcal{K}_1$, which is given as
\begin{equation}
\begin{split}
y^{j}_{k} =\mathbf{\tilde{g}}_{k,i}^H\mathbf{x}^{[1]}+\sum_{m\in\mathcal{K}_1}\underset{\tilde{h}_{m,k,i}}{\left(\underbrace{h_{m,k}+\mathbf{h}^{H}_{k}\mathbf{\Theta}_i^H\mathbf{h}_{m}} \right )}x^{[2]}_m+n_k,&\\
i\in\{r,t\}.&
\end{split}
\end{equation}
Each user in $\mathcal{K}_2$ then employs the  maximal ratio combining (MRC) approach to ensure the received signals are   properly co-phased and merged \cite{4257440}.  Denote the effective channel between user-$m$ and user-$n$ as $\tilde{h}_{m,n,i}$, the SINRs of decoding the $s_0$ and $s_k$  at  user-$k$ in $\mathcal{K}_2$ are given as
\begin{align}
\small
&\gamma^{\mathrm{FD}}_{c,k}=\underset{\gamma^{\mathrm{FD},[1]}_{c,k}}{\underbrace{\frac{\left|\mathbf{\tilde{g}}_{k,i}^H\mathbf{p}_{0}\right|^2}{\sum_{m\in\mathcal{K}}\left|\mathbf{\tilde{g}}_{k,i}^H\mathbf{p}_{m}\right|^2+\sigma^2}} }+\underset{\gamma^{\mathrm{FD},[2]}_{c,k}}{\underbrace{\frac{\sum_{m\in\mathcal{K}_1}\left|\tilde{h}_{m,k,i}\right|^2P_{m}}{\sigma^2}} },\label{Gamma_c2}\\
&\gamma^{\mathrm{FD}}_{k}=\frac{\left|\mathbf{\tilde{g}}_{k,i}^H\mathbf{p}_{k}\right|^2}{\sum_{m\in\mathcal{K},m\neq k}\left|\mathbf{\tilde{g}}_{k,i}^H\mathbf{p}_{m}\right|^2+\sigma^2}.\label{Gamma_p2}
\end{align}
With the SINRs of $s_0$ and $s_k$ in (\ref{Gamma_c1}), (\ref{Gamma_p1}), (\ref{Gamma_c2}), and (\ref{Gamma_p2}),  the achievable common  and private  rates are obtained as $R^{j}_{c,k}=\log_2(1+\gamma^{\mathrm{FD}}_{c,k})$ and $ R^{j}_{k}=\log_2(1+\gamma^{\mathrm{FD}}_{k})$, $j\in \{\mathrm{FE},\mathrm{FM}\}$.
To guarantee all users  decode the common stream, the following condition should satisfy \cite{mao2017rate}:
\begin{equation}
\label{R_c}
   R_c^{j}=\min\{R_{c,k}^{j}|k\in\mathcal{K}\},
\end{equation}
$R_c^{j}$ is shared by all users, we denote $C_k$ as the part of $R_c^{j}$ that is allocated to user-$k$, it satisfies $\sum_{k\in\mathcal{K}} C_k\leq R_c^{j}$.
Hence, the total achievable rate of user-$k$ is
\begin{equation}
\label{tot}
   R^{j}_{tot,k}=R^{j}_{k} +C_{k},\forall k \in \mathcal{K}. 
\end{equation}
Note that equations (\ref{R_c}) and (\ref{tot}) remain applicable for  other transmission modes by setting $j$ to $\mathrm{FT/HE/HM/HT}$.
 In the following discussion of these transmission modes, the equations  will not be repeated. 
\subsubsection{FT}
In FT mode, there are two time slots: all elements of STAR RIS operate in the reflection mode during the first time slot ($\lambda_r$) while all elements of STAR RIS operate in the transmission mode during the second time slot ($1-\lambda_r$).
We denote the SINRs for decoding $s_0$ and $s_k$ in the reflection and transmission time slots respectively as $\gamma^{\mathrm{FD},i}_{c,k}$ and $\gamma^{\mathrm{FD},i}_{k}$, where $i\in\{r,t\}$.
In the reflection time slot, 
users in  $\mathcal{K}_r$ receives signal from the BS and the STAR RIS while users in $\mathcal{K}_t$ receives signal from the BS only. The corresponding SINRs are given as  
\begin{equation}
    \begin{aligned}    \gamma^{\mathrm{FD},r}_{c,k}&=\gamma^{\mathrm{FD}}_{c,k}\mid_{\bm{\Theta}_i=\bm{\Theta}_r}, \,\,\gamma^{\mathrm{FD},r}_{k}=\gamma^{\mathrm{FD}}_{k}\mid_{\bm{\Theta}_i=\bm{\Theta}_r}, k\in \mathcal{K}_r,\\  \gamma^{\mathrm{FD},r}_{c,k}&=\gamma^{\mathrm{FD}}_{c,k}\mid_{\bm{\Theta}_i=\mathbf{0}}, \,\,\gamma^{\mathrm{FD},r}_{k}=\gamma^{\mathrm{FD}}_{k}\mid_{\bm{\Theta}_i=\mathbf{0}}, k\in \mathcal{K}_t,\\  
    \end{aligned}
\end{equation}
where $\gamma^{\mathrm{FD}}_{c,k}$ and $\gamma^{\mathrm{FD}}_{k}$ are defined in (\ref{Gamma_c1}) and (\ref{Gamma_p1}), respectively. 
In the transmission time slot, 
users in  $\mathcal{K}_r$ receives signal from the BS only while users in $\mathcal{K}_t$ receives signal from both BS and STAR RIS. The corresponding SINRs are given as  
\begin{equation}
    \begin{aligned}    \gamma^{\mathrm{FD},t}_{c,k}&=\gamma^{\mathrm{FD}}_{c,k}\mid_{\bm{\Theta}_i=\mathbf{0}}, \,\,\gamma^{\mathrm{FD},t}_{k}=\gamma^{\mathrm{FD}}_{k}\mid_{\bm{\Theta}_i=\mathbf{0}}, k\in \mathcal{K}_r,\\  \gamma^{\mathrm{FD},t}_{c,k}&=\gamma^{\mathrm{FD}}_{c,k}\mid_{\bm{\Theta}_i=\bm{\Theta}_t}, \,\,\gamma^{\mathrm{FD},t}_{k}=\gamma^{\mathrm{FD}}_{k}\mid_{\bm{\Theta}_i=\bm{\Theta}_t}, k\in \mathcal{K}_t\\  
    \end{aligned}
\end{equation}

The achievable common  and private rates for user-$k$ are given as 
\begin{equation}
\begin{split}    R^{\mathrm{FT}}_{c,k}&=\lambda_{r}\log_2(1+\gamma^{\mathrm{FD},r}_{c,k})+(1-\lambda_{r})\log_2(1+\gamma^{\mathrm{FD},t}_{c,k}),\\
R^{\mathrm{FT}}_{k}&=\lambda_{r}\log_2(1+\gamma^{\mathrm{FD},r}_{k})+(1-\lambda_{r})\log_2(1+\gamma^{\mathrm{FD},t}_{k}).
\end{split}
\end{equation}
By substituting $R^{\mathrm{FT}}_{c,k}$ and $R^{\mathrm{FT}}_{k}$  into (\ref{R_c}) and (\ref{tot}),  we obtain the corresponding achievable rate of each user in the FT mode. 
\subsection{Received Signal Model for HD-based Transmission Modes}
\label{CRS HD}
In HD-based transmission modes,  the direct and cooperative transmission phases  are orthogonal in each transmission block. 
Contrast to FD-based modes, there is no self-interference at  relaying users, but the transmission rates for the direct and transmission phases are scaled by the pre-log factors $\lambda (0 < \lambda \leq  1)$ and $(1-\lambda)$, respectively. In the following, we specify the received signals and achievable rates for HE, HM, and HT.
\subsubsection{HE/HM}
In the direct transmission phase, the  signal received at  user-$k$ is 
\begin{equation}
\begin{split}
\label{HD-signal}
  y^{\mathrm{HD},[1]}_{k}=\mathbf{\tilde{g}}^H_{k,i}\mathbf{x}^{[1]}+n_k,\forall k\in\mathcal{K},j\in \{\mathrm{HE}, \mathrm{HM}\}.
  \end{split}
  \end{equation}
The SINRs of decoding $s_0$ and $s_k$ in the direct transmission phase are given as
\begin{align}
   &\gamma^{\mathrm{HD},[1]}_{c,k}=\frac{\left|\mathbf{\tilde{g}}_{k,i}^H\mathbf{p}_{0}\right|^2}{\sum_{m\in\mathcal{K}}\left|\mathbf{\tilde{g}}_{k,i}^H\mathbf{p}_{m}\right|^2+\sigma^2},\label{HD:gamma_c1}\\     &\gamma^{\mathrm{HD}}_{k}=\frac{\left|\mathbf{\tilde{g}}_{k,i}^H\mathbf{p}_{0}\right|^2}{\sum_{m\in\mathcal{K},m\neq k}\left|\mathbf{\tilde{g}}_{k,i}^H\mathbf{p}_{m}\right|^2+\sigma^2}.\label{HD:gamma_p1}
\end{align}
In the cooperative transmission phase,  users in  $\mathcal{K}_1$ forward the decoded  $s_0$ to users in $\mathcal{K}_2$.
The  signal received at user-$k$ in  $\mathcal{K}_2$ in the cooperative transmission phase is 

\begin{align}
    y^{\mathrm{HD},[2]}_{k} =\sum_{m\in\mathcal{K}_1}\tilde{h}_{m,k,i}\sqrt{P_m}s_0+n_k,\forall k\in \mathcal{K}_2.
\end{align}
The SINR of decoding $s_0$ at  user-$k$ in $\mathcal{K}_2$ in the cooperative transmission phase is given as
\begin{align}
\label{HD:gamma_c2}
    \gamma^{\mathrm{HD},[2]}_{c,k}=\frac{\sum_{m\in\mathcal{K}_1}\left|\tilde{h}_{m,k,i}\right|^2P_{m}}{\sigma^2}.
\end{align}
With the SINR in (\ref{HD:gamma_c1}), (\ref{HD:gamma_p1}), and (\ref{HD:gamma_c2}),
we obtain the achievable common and private rates in HF and HM modes ($j\in\mathrm{HE}/\mathrm{HM}$) as
\begin{equation}
\label{eq:commonRateHEHM}
\small
\begin{aligned}
R^{j}_{c,k}&=\lambda\log_2\left(1+\gamma^{\mathrm{HD},[1]}_{c,k}\right), \forall k\in\mathcal{K}_1   \\
R^{j}_{c,k}&=\lambda\log_2\left(1+\gamma^{\mathrm{HD},[1]}_{c,k}\right)+(1-\lambda)\log_2\left(1+\gamma^{\mathrm{HD},[2]}_{c,k}\right), \forall k\in \mathcal{K}_2
\end{aligned}
\end{equation}
and
\begin{equation}
\label{eq:privateRateHEHM}
R^{j}_{k}=\lambda\log_2\left(1+\gamma^{\mathrm{HD}}_{k}\right).
\end{equation}
By substituting (\ref{eq:commonRateHEHM}) and (\ref{eq:privateRateHEHM})  into (\ref{R_c}) and (\ref{tot}),  we obtain the corresponding achievable rate of each user in the HE and HM modes. 
\subsubsection{HT}
In the HT mode, there are four time slots:
two for direct transmission with STAR RIS reflection or transmission, and two for cooperative transmission with STAR RIS reflection or transmission.
We use $\gamma^{\mathrm{HD},[s],i}_{c,k}$ and $\gamma^{\mathrm{HD},i}_{k}$, $i\in\{r,t\}$, $s\in\{1,2\}$ to denote the SINRs of decoding $s_0$ and $s_k$ in the direct or cooperative transmission phase. The superscript $i$ indicates whether it's in the reflection or transmission time slot.
Specifically, in the two reflection time slots respectively for cooperative and transmission phases, we have
\begin{equation}
    \begin{aligned}    \gamma^{\mathrm{HD},[s],r}_{c,k}&=\gamma^{\mathrm{HD},[s]}_{c,k}\mid_{\bm{\Theta}_i=\bm{\Theta}_r}, \,\,\gamma^{\mathrm{HD},r}_{k}=\gamma^{\mathrm{HD}}_{k}\mid_{\bm{\Theta}_i=\bm{\Theta}_r}, k\in \mathcal{K}_r,\\  \gamma^{\mathrm{HD},[s],r}_{c,k}&=\gamma^{\mathrm{HD},[s]}_{c,k}\mid_{\bm{\Theta}_i=\mathbf{0}}, \,\,\gamma^{\mathrm{HD},r}_{k}=\gamma^{\mathrm{HD}}_{k}\mid_{\bm{\Theta}_i=\mathbf{0}}, k\in \mathcal{K}_t,\\  
    \end{aligned}
\end{equation}
where $\gamma^{\mathrm{HD},[s]}_{c,k}, s\in\{1,2\}$ and $\gamma^{\mathrm{FD}}_{k}$ are defined in (\ref{HD:gamma_c1}), (\ref{HD:gamma_c2}), and (\ref{HD:gamma_p1}), and respectively. In the two transmission time slots respectively for cooperative and transmission phases, we have
\begin{equation}
    \begin{aligned}    \gamma^{\mathrm{HD},[s],t}_{c,k}&=\gamma^{\mathrm{HD},[s]}_{c,k}\mid_{\bm{\Theta}_i=\mathbf{0}}, \,\,\gamma^{\mathrm{HD},t}_{k}=\gamma^{\mathrm{HD}}_{k}\mid_{\bm{\Theta}_i=\mathbf{0}}, k\in \mathcal{K}_r,\\  \gamma^{\mathrm{HD},[s],t}_{c,k}&=\gamma^{\mathrm{HD},[s]}_{c,k}\mid_{\bm{\Theta}_i=\bm{\Theta}_t}, \,\,\gamma^{\mathrm{HD},t}_{k}=\gamma^{\mathrm{HD}}_{k}\mid_{\bm{\Theta}_i=\bm{\Theta}_t}, k\in \mathcal{K}_t.\\  
    \end{aligned}
\end{equation}

Hence, the achievable  common and private rates for user-$k$ are given as 
\begin{equation}
\label{HT:rate1}
\begin{aligned} R^{\mathrm{HT}}_{c,k}&=\lambda_1\log_2\left(1+\gamma^{\mathrm{HD},[1],r}_{c,k}\right)+\lambda_2\log_2\left(1+\gamma^{\mathrm{HD},[1],t}_{c,k}\right),\\
&\quad\quad\quad\quad\quad\quad\quad\quad\quad\quad\quad\quad\quad\quad\quad\quad\quad\forall k\in\mathcal{K}_1, \\
R^{\mathrm{HT}}_{c,k}&=\lambda_1\log_2\left(1+\gamma^{\mathrm{HD},[1],r}_{c,k}\right)
\\&+(\lambda_r-\lambda_1)\log_2\left(1+\gamma^{\mathrm{HD},[2],r}_{c,k}\right)\\
&+\lambda_2\log_2\left(1+\gamma^{\mathrm{HD},[1],t}_{c,k}\right)\\
&+(1-\lambda_r-\lambda_2)\log_2\left(1+\gamma^{\mathrm{HD},[2],t}_{c,k}\right),\quad\forall k\in\mathcal{K}_2.
\end{aligned}
\end{equation}

\begin{equation}
\label{HT:rate2}
R^{\mathrm{HT}}_{k}=\lambda_1\log_2\left(1+\gamma^{\mathrm{HD},r}_{k}\right)+\lambda_2\log_2\left(1+\gamma^{\mathrm{HD},t}_{k}\right).
\end{equation}


\subsection{Problem Formulation}
In this work, we aim to jointly optimize the active beamforming  $\mathbf{P}$ at the BS,  the common rate allocation $\mathbf{c} = [C_1,\cdots,C_K]$, as well as the passive beamforming  $\bm{\Theta}_r$, $\bm{\Theta}_t$ at the STAR RIS. For the HD-based transmission modes, the corresponding time allocation coefficients are also jointly optimized. Our objective is to maximize the minimum user 
rate (max-min rate) among all users subject to the transmit power constraint. 
The formulated  problems for the six transmission modes of the proposed STAR RIS-aided CRS model are illustrated as follows: 
\subsubsection{FE}
$j=\mathrm{FE}$, the optimization problem is formulated as 
\begin{subequations}
\label{FD-ES}
\begin{align}
\mathcal{P}_{\mathrm{FE}}:\,\,\,\,&\max_{\mathbf{P}, \mathbf{c}, \bm{\Theta}_{r},\bm{\Theta}_{t}             }\,\,\min_{k\in \mathcal{K}} \,\,R^{\mathrm{FE}}_{tot,k} \label{FD-ES:1}\\
\mbox{s.t.}\quad
	& \sum_{k\in\mathcal{K}}C_{k}\leq R^{j}_{c}, C_{k}\geq 0,\forall k\in \mathcal{K} \label{FD-ES:2}\\
	&\theta_{n,i}\in [0,2\pi) ,\forall n\in \mathcal{N},i\in \{r,t\}, \label{FD-ES:4}\\
	&\text{tr}\left(\mathbf{P}\mathbf{P}^{H}\right)\leq P_{t} \label{FD-ES:5}\\ &\beta_{n,r},\beta_{n,t}\in[0,1],\beta_{n,r}+\beta_{n,t}=1,\forall n\in\mathcal{N},\label{FD-ES:6}
\end{align}
\end{subequations}
where constraint (\ref{FD-ES:2}) guarantees  the common stream is successfully decoded by all users. 
Constraints (\ref{FD-ES:4}) and (\ref{FD-ES:6}) are the  constraints for the phase shift and amplitude ranges for each element of the STAR RIS, respectively.
(\ref{FD-ES:5}) limits the  transmit power at the BS.
\subsubsection{FM}
$j=\mathrm{FM}$,  the formulated problem is given as
\begin{subequations}
  \label{FD-MS}
  \begin{align}
    \mathcal{P}_{\mathrm{FM}}:\,\,\,\,&\max_{\mathbf{P}, \mathbf{c}, \bm{\Theta}_{r},\bm{\Theta}_{t}             }\,\,\min_{k\in \mathcal{K}} \,\,R^{\mathrm{FM}}_{tot,k} \label{FD-MS:1}\\
\mbox{s.t.}\quad
	& \text{(\ref{FD-ES:2})--(\ref{FD-ES:5})} \nonumber\\
	&\beta_{n,r},\beta_{n,t}\in\{0,1\},\beta_{n,r}+\beta_{n,t}=1,\forall n\in\mathcal{N}. \label{FD-MS:3}   \end{align}
\end{subequations}

\subsubsection{FT}
$j=\mathrm{FT}$,  the time allocation variable $\lambda_r$   is  jointly optimized with the beamforming matrices and the common rate allocation. The optimization problem is 
\begin{subequations}
\label{FD-TS}
\begin{align}
\mathcal{P}_{\mathrm{FT}}:\,\,\,\,&\max_{\mathbf{P}, \mathbf{c}, \bm{\Theta}_{r},\bm{\Theta}_{t},\lambda_r             }\,\,\min_{k\in \mathcal{K}} \,\,R^{\mathrm{FT}}_{tot,k} \label{FD-TS:1}\\
\mbox{s.t.}\quad
	& \text{(\ref{FD-ES:2})--(\ref{FD-ES:5}),}  \nonumber\\
 &\beta_{n,r},\beta_{n,t}\in[0,1],\forall n\in\mathcal{N},\label{FD-TS:2}\\
 & 0\leq \lambda_r \leq 1. \label{FD-TS:3}
\end{align}
\end{subequations}
\subsubsection{HE}
$j=\mathrm{HE}$, the time allocation variable  $\lambda$  is jointly optimized with the beamforming matrices and the common rate allocation. The formulated problem is 
\begin{subequations}
\label{HD-ES}
\begin{align}
\mathcal{P}_{\mathrm{HE}}:\,\,\,\,&\max_{\mathbf{P}, \mathbf{c}, \bm{\Theta}_{r},\bm{\Theta}_{t},\lambda }\,\,\min_{k\in \mathcal{K}} \,\,R^{\mathrm{HE}}_{tot,k} \label{HD-ES:1}\\
\mbox{s.t.}\quad
	& \text{(\ref{FD-ES:2})--(\ref{FD-ES:6}) }  \nonumber\\
 & 0< \lambda \leq 1. \label{HD-ES:2}
\end{align}
\end{subequations}
\subsubsection{HM}
 $j=\mathrm{HM}$,  the corresponding  optimization problem is formulated as 
\begin{subequations}
  \label{HD-MS}
  \begin{align}    \mathcal{P}_{\mathrm{HM}}:\,\,\,\,&\max_{\mathbf{P}, \mathbf{c}, \bm{\Theta}_{r},\bm{\Theta}_{t},\lambda             }\,\,\min_{k\in \mathcal{K}} \,\,R^{\mathrm{HM}}_{tot,k} \label{HD-MS:1}\\
\mbox{s.t.}\quad
	& \text{(\ref{FD-ES:2})--(\ref{FD-ES:5}),
 (\ref{FD-MS:3}), (\ref{HD-ES:2}).} \nonumber
  \end{align}
\end{subequations}
\subsubsection{HT}
$j=\mathrm{HT}$, $\lambda_r, \lambda_1, \lambda_2$ are jointly optimized. 
The corresponding optimization problem is formulated as
\begin{subequations}
\label{HD-TS}
\begin{align}
\mathcal{P}_{\mathrm{HT}}:\,\,\,\,&\max_{\mathbf{P}, \mathbf{c}, \bm{\Theta}_{r},\bm{\Theta}_{t}, \lambda_r,\lambda_1,\lambda_2}\,\,\min_{k\in \mathcal{K}} \,\,R^{\mathrm{HT}}_{tot,k} \label{HD-TS:1}\\
\mbox{s.t.}\quad
	& \text{(\ref{FD-ES:2})--(\ref{FD-ES:5}), (\ref{FD-TS:2}), (\ref{FD-TS:3})}  \nonumber\\
 & 0< \lambda_1 \leq \lambda_r, 0< \lambda_2 \leq 1- \lambda_r. \label{HD-TS:2}
\end{align}
\end{subequations}
From  problems (\ref{FD-ES})-(\ref{HD-TS})  for the six transmission modes,  we observe that 
$\mathbf{P}$, $\bm{\Theta}_r$, $\bm{\Theta}_t$ are highly coupled in the non-convex and fractional expressions for the  achievable common and private rates. 
Besides, problem (\ref{FD-MS}) and (\ref{HD-MS}) involve integer programming due to (\ref{FD-ES:2}).  All  problems are highly non-convex and   difficult to solve.
In the next section,  we will propose two algorithm to solve the problems.

\section{Proposed Optimization Frameworks}
\label{optimization framework}
To solve the aforementioned six non-convex optimization problems, in this section, we first propose an AO framework that alternatively optimizes the STAR RIS passive beamforming matrices $\bm{\Theta}_r$ and $\bm{\Theta}_t$, and the remaining variables. To ease the computational complexity, we further propose a low-complexity algorithm to address the optimization problems. 
\subsection{Proposed AO Algorithm}
\label{AO}
We divide the original non-convex problems into two subproblems, one for the STAR RIS passive beamforming design, the other for the joint BS active beamforming, common rate, and time allocation design. The approaches for solving the two subproblems are detailed in the following:
\subsubsection{STAR RIS passive beamforming optimization}
With the transmit beamforming $\mathbf{P}$, common rate allocation $\mathbf{c}$ and time allocation variables $\lambda_t$, $\lambda_1$ and $\lambda_2$  fixed, we focus on optimizing $\bm{\Theta}_r$ and $\bm{\Theta}_t$.
First, we introduce an auxiliary variable $t^j$ to represent the objective function for the transmission mode $j$,  leading to the following constraint:  
\begin{equation}
\label{objective}
    R^{j}_{p,k}+C_{k}\geq t^j,\forall k\in\mathcal{K},j\in\{\mathrm{FE},\mathrm{FM},\mathrm{FT},\mathrm{HE},\mathrm{HM},\mathrm{HT}\}.
\end{equation}  

\par 
Although the achievable rate expressions may vary across different transmission modes, our main focus is the SINR expressions as the optimization variables $\bm{\Theta}_r$ and $\bm{\Theta}_t$ are embedded solely within the SINR. Note that the non-convexity of the SINR expressions remains consistent across all six transmission modes. The key distinction in optimizing the STAR RIS across these modes lies in the constraints imposed by the ES, MS, and TS protocols. 
\par 
In the following, we begin by illustrating the SCA-based method employed to address the non-convexity of the SINR expressions, using the FE mode as an example. Following this, we elaborate the penalty methods proposed to manage the STAR RIS constraints for the ES, MS, and TS protocols, respectively. 

\begin{itemize}
    \item   \textbf{Step 1.1: The SCA method for non-convex SINRs}:
\end{itemize}
\par 
Let $\psi_{n,i}=\sqrt{\beta_{n,i}}e^{j\theta_{n,i}},n\in\mathcal{N},i\in\{r,t\}$ denote each element of $\bm{\Theta}_r$ and $\bm{\Theta}_t$.
With $\bm{\psi}_i=[\psi_{1,i},\ldots,\psi_{N,i}]^T$, we have $\mathbf{g}_k^H\mathbf{p}_{m}+\mathbf{h}_{k}^H\mathbf{\Theta}_{i}\mathbf{E}\mathbf{p}_{m}=\bar{g}_{km}+\mathbf{s}_{km}^{H}\bm{\psi}_{i}$, $\mathbf{h}_{n}^H\mathbf{\Theta}_{i}\mathbf{h}_{m}=\bar{\mathbf{h}}_{m,n}^{H}\bm{\psi}_{i}$, where $\bar{g}_{km}=\mathbf{g}_k^H\mathbf{p}_{m}$, $\mathbf{s}_{km}=(\textrm{diag}(\mathbf{h}_{k}^{H})\mathbf{E}\mathbf{p}_{m})^{\ast}$ and $\bar{\mathbf{h}}_{m,n}=(\textrm{diag}(\mathbf{h}_{n}^{H})\mathbf{h}_{m})^{\ast}\in \mathbb{C}^{N\times 1}$. In this way, the optimization variables are represented by  $\bm{\psi}_{r}$ and $\bm{\psi}_{t}$.
\par 
Taking the FE problem (\ref{FD-ES}) as an example, we first introduce slack variables $\bm{\delta}=[\delta_1,\ldots,\delta_K]^T$ and $\bm{\delta}_c=[\delta_{c,1},\ldots,\delta_{c,K}]^T$ to respectively denote the SINRs of $s_k$ and $s_0$ in the direct transmission phase.
Furthermore, we introduce  $\bm{\xi}=[\xi_1,\ldots,\xi_{K_2}]$ to denote the  SINRs of the desination users in the cooperative transmission phase, where $K_2$ is the  number of  users in $\mathcal{K}_2$.
The optimization problem (\ref{FD-ES}) for updating the STAR RIS passive beamforming matrices is then equivalently transformed into
\begin{subequations}
\label{Rate}
    \begin{align}
      \mathcal{P}_{\mathrm{FE-1:}}\,\,\,\,&\max_{\substack{t^{\mathrm{FE}},\bm{\psi}_{r},\bm{\psi}_{t},
      \bm{\delta},\bm{\delta}_c, \bm{\xi} }            }\,\,t^{\mathrm{FE}}\\  
\mbox{s.t.}\quad    	
     & \log_2(1+\delta_k)+C_{k}\geq t^{\mathrm{FE}},\forall k\in\mathcal{K}, \label{Rate:1}\\
    & \log_2(1+\delta_{c,k}) \geq {\textstyle \sum_{m\in\mathcal{K}}}C_m,k\in\mathcal{K}_1, \label{Rate:2}\\
    & \log_2(1+\delta_{c,k}+\xi_k) \geq {\textstyle \sum_{m\in\mathcal{K}}}C_m,k\in\mathcal{K}_2, \label{Rate:3}\\
    &\delta_{c,k}\leq \gamma^{\mathrm{FD},[1]}_{c,k},\forall k\in\mathcal{K},\label{Rate:4}\\    
    &\delta_{k}\leq \gamma^{\mathrm{FD}}_{k},\forall k\in\mathcal{K},\label{Rate:5}\\
     &\xi_{k}\leq \gamma^{\mathrm{FD},[2]}_{c,k},\forall k\in\mathcal{K}_2,\label{Rate:6}\\
     &\text{(\ref{FD-ES:6})}.\nonumber
    \end{align}
\end{subequations}
For  constraints (\ref{Rate:4}) and (\ref{Rate:5}), we further introduce slack variables $\bm{\eta}=[\eta_1,\ldots,\eta_K]^T$ and $\bm{\eta}_c=[\eta_{c,1},\ldots,\eta_{c,K}]^T$  to respectively denote the  denominators of the SINRs for $s_0$ and $s_k$.   (\ref{Rate:4}) and (\ref{Rate:5}) are equivalently transformed to 
\begin{subequations}
\label{denominator}
\label{eq:step11}
    \begin{align}
   & \eta_{c,k}\geq \mathrm{de}(\gamma^{\mathrm{FD},[1]}_{c,k}),\forall k\in\mathcal{K}\\
       & \eta_k\geq \mathrm{de}(\gamma^{\mathrm{FD}}_{k}),\forall k\in\mathcal{K}\\
        &\delta_{c,k}\eta_{c,k}\leq \left|\bar{g}_{k0}+\mathbf{s}_{k0}^{H}\bm{\psi}_{i}\right|^2,\label{Fe:1}\\
        &\delta_k\eta_k\leq \left|\bar{g}_{kk}+\mathbf{s}_{kk}^{H}\bm{\psi}_{i}\right|^2,\label{Fe:2}
    \end{align}
\end{subequations}
where $\mathrm{de}(\cdot)$ is an operator  defined to extract the denominator of a fraction, i.e., $\mathrm{de}(\frac{a}{b})=b$.
For   constraint (\ref{Fe:1}), the left-hand side (LHS) equals to $\delta_{c,k}\eta_{c,k}=\frac{1}{4}(\delta_{c,k}+\eta_{c,k})^2-\frac{1}{4}(\delta_{c,k}-\eta_{c,k})^2$. 
By further approximating $\delta_{c,k}\eta_{c,k}$ at the point $(\delta_{c,k}^{[z]}, \eta_{c,k}^{[z]})$ in iteration $[z]$ using the first-order Taylor approximation of $(\delta_{c,k}-\eta_{c,k})^2$, we have 
\begin{equation}
\label{denominator:1}
    \begin{split}
        \delta_{c,k}\eta_{c,k}\leq &\frac{1}{4}(\delta_{c,k}+\eta_{c,k})^2-\frac{1}{2}(\delta_{c,k}^{[z]}-\eta_{c,k}^{[z]})(\delta_{c,k}-\eta_{c,k})\\
        &+\frac{1}{4}(\delta_{c,k}^{[z]}-\eta_{c,k}^{[z]})^2 \triangleq 
 \nu(\delta_{c,k}^{[z]},\eta_{c,k}^{[z]}).
    \end{split}
\end{equation}
We also approximate the right-hand side (RHS) of (\ref{Fe:1}) at $\bm{\psi}^{[z]}_i$ by the first-order Taylor approximation of $\left|\bar{g}_{k0}+\mathbf{s}_{k0}^{H}\bm{\psi}_{i}\right|^2$ as
\begin{equation}
\label{numerator}
    \begin{split}
\left|\bar{g}_{k0}+\mathbf{s}_{k0}^{H}\bm{\psi}_{i}\right|^2\geq &2\Re\{(\mathbf{s}^H_{k0}\bm{\psi}^{[z]}_i+\bar{g}_{k0})^H\mathbf{s}^H_{k0}\bm{\psi}_i\}-|\mathbf{s}^H_{k0}\bm{\psi}_i|^2\\
&+|\bar{g}_{k0}|^2\triangleq \varpi(\bm{\psi}_i^{[z]},\bm{\psi}_i,\bar{g}_{k0},\mathbf{s}_{k0}).
    \end{split}
\end{equation}
Based on this approach,  constraints (\ref{Fe:1}) and (\ref{Fe:2}) are transformed to  
\begin{subequations}
\label{numerator-denominator}
    \begin{align}
        &\nu(\delta_{c,k}^{[z]},\eta_{c,k}^{[z]})\leq \varpi(\bm{\psi}_i^{[z]},\bm{\psi}_i,\bar{g}_{k0},\mathbf{s}_{k0}),\forall k\in\mathcal{K}_1, \label{numerator-denominator:1}\\ 
        &\nu(\delta_{k}^{[z]},\eta_{k}^{[z]})\leq \varpi(\bm{\psi}_i^{[z]},\bm{\psi}_i,\bar{g}_{kk},\mathbf{s}_{kk}),\forall k\in\mathcal{K}.  \label{numerator-denominator:3}
    \end{align}
\end{subequations}
And constraint (\ref{Rate:6}) is approximated as
\begin{equation}
\label{FD:SINR}
    \xi_k\sigma^2 \leq P_m\varpi(\bm{\psi}_i^{[z]},\bm{\psi}_i,h_{m,k},\mathbf{h}_{m,k}),\forall k\in\mathcal{K}
\end{equation}
\par Therefore, utilizing the SCA method, the non-convex SINR expressions  (\ref{Rate:4})--(\ref{Rate:6}) are replaced by (\ref{numerator-denominator}) and (\ref{FD:SINR}), which are convex. This approach can be applied directly to  other five transmission modes, so we simplify by omitting their discussion here.

\begin{itemize}
    \item   \textbf{Step 1.2: The penalty method for non-convex STAR RIS constraints}:
\end{itemize}
After Step 1.1, the optimization problems for passive beamforming in all six modes become convex except for the constraints related to the phase and amplitude of the STAR RIS. In the following, we specify the penalty methods proposed to handle these constraints for the ES, MS, and TS protocols.

\textit{ES}:
For constraint (\ref{FD-ES:6}), we adopt the penalty method proposed in \cite{yang2021energy}. By introducing a large positive constant $C$, the objective function into $t^{j}+C\sum_{n=1}^N(|\psi_{n,r}|^2+|\psi_{n,t}|^2-1), j\in\{\mathrm{FE},\mathrm{HE}\}$
with an additional constraint
\begin{equation}
\label{appro:theta}
    |\psi_{n,r}|^2+|\psi_{n,t}|^2\leq 1.
\end{equation}
Using  the first-order Taylor approximation of $(|\psi_{n,r}|^2+|\psi_{n,t}|^2)$ at iteration $[z]$, we obtain the approximated objective function as
 \begin{equation}
 \label{obj:2}
     t^{j}+C\sum_{n=1}^N\Re \{2(\psi^{[z]}_{n,r})^{\ast}\psi_{n,r}-|\psi_{n,r}^{[z]}|^2+2(\psi^{[z]}_{n,t})^{\ast}\psi_{n,t}-|\psi_{n,t}^{[z]}|^2\}.
 \end{equation}
\par 
\textit{MS}: 
For constraint (\ref{FD-MS:3}), it is obvious that the amplitude coefficient of each STAR RIS element in the  FM and HM modes is an integer chosen from 0 and 1. To transform constraints (\ref{FD-MS:3}), it's worth noting that no matter each element is 0 or 1, it always satisfies (\ref{appro:theta}) and
\begin{equation}
\label{element:FM}
    -|\psi_{n,i}|^2+|\psi_{n,i}|=0, i\in\{r,t\} .
\end{equation}
(\ref{element:FM})  forces the amplitudes of $\psi_{n,r}$ and $\psi_{n,t}$ to 0 or 1.
Due to constraint (\ref{appro:theta}), $ 0\leq|\psi_{n,i}|\leq1$, we obtain that
\begin{equation}
\label{FM:theta2}
    -|\psi_{n,i}|^2+|\psi_{n,i}|\geq 0,i\in\{r,t\}.
\end{equation}
Hence, constraint (\ref{FD-MS:3}) can be replaced by  adding a penalty term  to the objective function  as $t^{j}-C\sum^{N}_{n=1}(-|\psi_{n,r}|^2+|\psi_{n,r}|-|\psi_{n,t}|^2+|\psi_{n,t}|),j\in \{\mathrm{FM},\mathrm{HM}\}$.
Applying the first-order Taylor approximation to the penalty term $-|\psi_{n,r}|^2-|\psi_{n,t}|^2$  at iteration $[z]$, the objective function is transformed to
\begin{equation}
\label{FM:objective}
    t^{j}-C\sum^{N}_{n=1}\left(A^{[z]}-2\Re\{(\psi_{n,r}^{[z]})^*\psi_{n,r}\}-2\Re\{(\psi_{n,t}^{[z]})^*\psi_{n,t}\}\right),
\end{equation}
where $A^{[z]}=|\psi_{n,r}^{[z]}|^2+|\psi_{n,t}^{[z]}|^2+|\psi_{n,r}|+|\psi_{n,t}|$.
\par
\textit{TS}: 
For TS modes, we also adopt the penalty method proposed in \cite{yang2021energy} to address constraints (\ref{FD-TS:2}), which transforms the objective fucntion into  $t^{j}+C\sum_{n=1}^{N}(|\psi_{n,r}|^2-1+|\psi_{n,t}|^2-1),j\in\{\mathrm{FT},\mathrm{HT}\}$
with an additional constraint
\begin{equation}
\label{FT:phase}
    |\psi_{n,i}|^2\leq 1, n\in\mathcal{N},i\in\{r,t\}.
\end{equation}
The objective function is then approximated in the same way as (\ref{obj:2}).
\par 
After Steps 1.1 and 1.2, the STAR RIS optimization subproblems for all six transmission modes become convex quadratically constrained quadratic program (QCQP) and  can be solved using the CVX optimization toolbox.
The detailed process to solve the STAR RIS passive beamforming optimization problem is  summarized in Algorithm \ref{Algorithm_1}.
\begin{algorithm}[!t]	
\label{Algorithm_1}
\textbf{Initialize}: $j$, $z=0$, $t^{j,[z]}$, $\bm{\Theta}^{[z]}_r$, $\bm{\Theta}^{[z]}_t$,  and slack variables for mode $j$\; 
 	\Repeat{$\text{convergence}$}{
 		$z\leftarrow z+1$\;
 Updating $\bm{\Theta}_r^{[z]},\bm{\Theta}_t^{[z]}$ by employing Steps 1.1 and 1.2 to approximate the problem and then solving the approximated convex problem using CVX \;			 	}	
\caption{ STAR RIS passive beamforming optimization algorithm for six transmission modes}
\end{algorithm}

\subsubsection{Joint optimization of  the active beamforming, common rate and time allocation}
Given  $\bm{\Theta}_r$ and $\bm{\Theta}_t$, the effective channels of all users are fixed. To jointly optimize the remaining variables, we first introduce an auxiliary variable $x^j$ to denote the objective function for the transmission mode $j$,  leading to the following constraints:  
\begin{equation}
\label{objective:P}
    R^{j}_{p,k}+C_{k}\geq x^j,\forall k\in\mathcal{K},j\in\{\mathrm{FE},\mathrm{FM},\mathrm{FT},\mathrm{HE},\mathrm{HM},\mathrm{HT}\}.
\end{equation}
In this subproblem,   non-convexity arises from both the fractional SINR expressions and the coupling of time allocation variables within the rate expressions. Both of these challenges are tackled using SCA in the following steps.

\begin{itemize}
    \item   \textbf{Step 2.1: The SCA method for time allocation variables}:
\end{itemize}
\par
In this step, we first specified the SCA methods employed to address the coupled time allocation variables within the rate expressions. Note that in the FE and FM modes, there are no time allocation variables. The non-convexity of the corresponding optimization problems arises solely from the SINR expressions, which are directly addressed in Step 2.2. 


\textit{FT}:
To optimize $\lambda_r$ in problem (\ref{FD-TS}), we first introduce the slack variables  $\bm{\alpha}^i=[\alpha^i_{1},\ldots,\alpha^i_{K}]^T$, and $\bm{\alpha}^i_c=[\alpha^i_{c,1},\ldots,\alpha^i_{c,K}]^T$, $i\in\{r,t\}$ to denote  the private and common rates. This allows us to transform  (\ref{objective:P}) and (\ref{FD-ES:2}) to
\begin{subequations}
\label{P_FT:objective2}
    \begin{align}
&\lambda_r\alpha^r_k+(1-\lambda_r)\alpha^t_k+C_{k}\geq x^{\mathrm{HT}}, \label{P_FT:Rate1}\\
& \lambda_r\alpha^r_{c,k}+(1-\lambda_r)\alpha^t_{c,k} \geq {\textstyle \sum_{m\in\mathcal{K}}}C_m,\label{P_FT:Rate2}\\
& \alpha^i_k\leq \log_2(1+\gamma^{\mathrm{FD},i}_k),\label{P_FT:Rate3}\\
& \alpha^i_{c,k}\leq \log_2(1+\gamma^{\mathrm{FD},i}_{c,k}),\label{P_FT:Rate4}.
\end{align}
\end{subequations}
For the bilinear function $\lambda_r\alpha^r_k$, it can be approximated by the first-order Taylor approximation at iteration $[z]$ based on $\nu(\cdot)$  in (\ref{denominator:1}).
Hence, constraints (\ref{P_FT:Rate1}) and (\ref{P_FT:Rate2}) are approximated as 
\begin{subequations}
\label{P_FT:bilinear}
    \begin{align}
        &-\nu(\alpha^{r,[z]}_k,-\lambda_r)+\alpha^{t}_k-\nu(\alpha^{t,[z]}_k,-\lambda_r)+C_k\geq x^{\mathrm{HT}},\label{P_FT:bilinear_1}\\
        &-\nu(\alpha^{r,[z]}_{c,k},-\lambda_r)+\alpha^{t}_{c,k}-\nu(\alpha^{t,[z]}_{c,k},-\lambda_r)\geq {\textstyle \sum_{m\in\mathcal{K}}}C_m.\label{P_FT:bilinear_2}
    \end{align}
\end{subequations}

\textit{HE/HM}: To optimize $\lambda$ in problems (\ref{HD-ES}) and (\ref{HD-MS}), akin to the FT mode, we introduce the slack variables  $\bm{\alpha}$, and $\bm{\alpha}^i_c$ to represent  the private and common rates, thereby facilitating the transformation of  (\ref{objective:P}) and (\ref{FD-ES:2}) to
\begin{subequations}
\label{P_HE:objective2}
    \begin{align}
&\lambda\alpha_k+C_{k}\geq x^{j}, \label{P_HE:Rate1}\\
& \lambda\alpha_{c,k} \geq {\textstyle \sum_{m\in\mathcal{K}}}C_m,k\in \mathcal{K}_1\label{P_HE:Rate2}\\
& \lambda\alpha_{c,k}+(1-\lambda)\log(1+\gamma^{\mathrm{HD},[2]}_{c,k})  \geq {\textstyle \sum_{m\in\mathcal{K}}}C_m,k\in \mathcal{K}_2\label{P_HE:Rate3}\\
& \alpha_k\leq \log_2(1+\gamma^{\mathrm{HD}}_k),\label{P_HE:Rate4}\\
& \alpha_{c,k}\leq \log_2(1+\gamma^{\mathrm{HD},[1]}_{c,k}),\label{P_HE:Rate5}.
\end{align}
\end{subequations}
Using the same first-order Taylor approximation method in (\ref{P_FT:bilinear}), constraints (\ref{P_HE:Rate1})--(\ref{P_HE:Rate3}) are approximated to 
\begin{align}
    &-\nu(\alpha^{[z]}_k,-\lambda)+C_{k}\geq x^{j}, \label{P_HE:Rate6}\\
& -\nu(\alpha^{[z]}_{c,k},-\lambda) \geq {\textstyle \sum_{m\in\mathcal{K}}}C_m,\forall k\in\mathcal{K}_1,\label{P_HE:Rate7}\\
& -\nu(\alpha^{[z]}_{c,k},-\lambda)+(1-\lambda) \log_2(1+\gamma^{\mathrm{HD},[2]}_{c,k}) \geq {\textstyle \sum_{m\in\mathcal{K}}}C_m,\nonumber\\
&\forall k\in\mathcal{K}_2.\label{P_HE:Rate8}
\end{align}

\textit{HT}:
To jointly optimize $\lambda_r$, $\lambda_1$, and $\lambda_2$ in problems (\ref{HD-ES}), we also introduce the slack variables  $\bm{\alpha}^i$, and $\bm{\alpha}^i_c$, $i\in\{r,t\}$ to  transform the problem. By applying the approximation method described in (\ref{P_FT:bilinear}), we approximate (\ref{objective:P}) and (\ref{FD-ES:2}) to
\begin{subequations}
\label{P_HT:objective2}
    \begin{align}
&-\nu(\alpha^{r,[z]}_k,-\lambda_1)-\nu(\alpha^{t,[z]}_k,-\lambda_2)+C_{k}\geq x^{\mathrm{FT}}, \label{P_HT:Rate1}\\
& -\nu(\alpha^{r,[z]}_{c,k},-\lambda_1) -\nu(\alpha^{t,[z]}_{c,k},-\lambda_2)\geq {\textstyle \sum_{m\in\mathcal{K}}}C_m,\forall k\in\mathcal{K}_1,\label{P_HT:Rate2}\\
& -\nu(\alpha^{r,[z]}_{c,k},-\lambda_1)+(\lambda_r-\lambda_1)\log_2(1+\gamma^{\mathrm{HD},[2],r}_{c,k})-\nonumber\\
&\nu(\alpha^{r,[z]}_{c,k},-\lambda_2)
+(1-\lambda_r-\lambda_2)\log_2(1+\gamma^{\mathrm{HD},[2],t}_{c,k}) \nonumber\\
&\geq {\textstyle \sum_{m\in\mathcal{K}}}C_m,\forall k\in\mathcal{K}_2,\label{P_HT:Rate3}\\
& \alpha^i_{c,k}\geq \log(1+\gamma^{\mathrm{HD},[1],i}_{c,k}),\label{P_HT:Rate6}\\
& \alpha^i_k\geq \log(1+\gamma^{\mathrm{HD},i}_{k}).\label{P_HT:Rate7}
\end{align}
\end{subequations}
Following Step 2.1, the time allocation variables are decoupled from the non-convex SINR expressions for all transmission modes. Next, we proceed to address the classical non-convex SINR expressions.

\begin{itemize}
    \item   \textbf{Step 2.2: The SCA method for non-convex SINRs}:
\end{itemize}
\par
Similar to Step 1.1, the SCA method for SINRs is applicable to all transmission modes. Therefore, we illustrate this approach using the FE problem (\ref{FD-ES}) as an example, without detailing other modes.

We first introduce slack variables $\bm{\iota}=[\iota_1,\cdots,\iota_K]^T$, $\bm{\iota}_c=[\iota_{c,1},\cdots,\iota_{c,K}]^T$ to denote the SINRs  of $s_k$ and $s_0$.
Constraints (\ref{objective:P}) and (\ref{FD-ES:2})  are equivalently transformed to 
\begin{subequations}
\label{P_FE:Rate}
    \begin{align}
        &\log_2(1+\iota_k)+C_{k}\geq x^{\mathrm{FE}},\forall k\in\mathcal{K},\label{P_FE:Rate_1}\\
        & \log_2(1+\iota_{c,k}) \geq {\textstyle \sum_{m\in\mathcal{K}}}C_m,k\in\mathcal{K}, \label{P_FE:Rate_2}\\
        &\iota_{c,k}\leq \gamma^{\mathrm{FD},[1]}_{c,k},\forall k\in\mathcal{K},\label{P_FE:Rate_3}\\
        &\iota_{k}\leq \gamma^{\mathrm{FD}}_{k},\forall k\in\mathcal{K}.\label{P_FE:Rate_4}
    \end{align}
\end{subequations}
Constraints (\ref{P_FE:Rate_3}) and (\ref{P_FE:Rate_4}) remain non-convex, we then introduce variables $\bm{\zeta}=[\zeta_1,\cdots,\zeta_K]^T$, $\bm{\zeta}_c=[\zeta_{c,1},\cdots,\zeta_{c,K}]^T$ to respectively denote the denominators of SINRs for $s_k$ and $s_0$. 
 (\ref{P_FE:Rate_3}) and (\ref{P_FE:Rate_4}) are equivalently transformed to 
\begin{subequations}
\label{P_FE:denominator}
    \begin{align}
   & \zeta_{c,k}\geq \mathrm{de}(\gamma^{\mathrm{FD},[1]}_{c,k}),\forall k\in\mathcal{K},\label{P_FE:denominator_1}\\
& \zeta_k\geq \mathrm{de}(\gamma^{\mathrm{FD}}_{k}),\forall k\in\mathcal{K}, \label{P_FE:denominator_2}\\
&\iota_{c,k}\zeta_{c,k}\leq \left|\tilde{\mathbf{g}}^H_{i,k}\mathbf{p}_{0}\right|^2,\label{P_FE:denominator_3}\\
&\iota_k\zeta_k\leq \left|\tilde{\mathbf{g}}^H_{i,k}\mathbf{p}_{k}\right|^2.\label{P_FE:denominator_4}
\end{align}
\end{subequations}
Following the approaches specified in Step 1.1 to address the non-convexity in (\ref{eq:step11}), we approximate constraints (\ref{P_FE:denominator_3}) and (\ref{P_FE:denominator_4}) at iteration $[z]$ to 
\begin{subequations}
\label{P_FE:numerator-denominator}
    \begin{align}
        &\nu(\iota_{c,k}^{[z]},\zeta_{c,k}^{[z]})\leq \varpi(\mathbf{p}_0^{[z]},\mathbf{p}_0,0,\tilde{\mathbf{g}}_{i,k}),\forall k\in\mathcal{K}, \label{P_FE:numerator-denominator_1}\\  
        &\nu(\iota_{k}^{[z]},\zeta_{k}^{[z]})\leq \varpi(\mathbf{p}_k^{[z]},\mathbf{p}_k,0,\tilde{\mathbf{g}}_{i,k}),\forall k\in\mathcal{K}.  \label{P_FE:numerator-denominator_3}
    \end{align}
\end{subequations}
\par
Based on the approximation methods specified in Step 2.1 and 2.2, the joint active beamforming, common rate, and time allocation optimization problems for all six transmission modes become QCQP and  can  be solved using CVX.
The detailed process of the SCA methods to solve this subproblem is illustrated in Algorithm \ref{Algorithm_2}.

\begin{algorithm}[!t]	
\label{Algorithm_2}
 	\textbf{Initialize}: $j, z=0, x^{j,[z]},\mathbf{P}^{[z]},\mathbf{c}^{[z]},\lambda_r^{[z]},\lambda^{[z]},\lambda_1^{[z]},\lambda_2^{[z]}$ and slack variables according to corresponding mode $j$\;
 	\Repeat{\textit{convergence}}{
 		$z\leftarrow z+1$\;
 Updating $\mathbf{P}^{[z]},\mathbf{c}^{[z]},\lambda_r^{[z]},\lambda^{[z]},\lambda_1^{[z]},\lambda_2^{[z]}$ uses  the SCA methods in Steps 2.1 and  2.2\;			 	}	
\caption{Joint transmit active beamforming and resource allocation algorithm }
\end{algorithm}
\par 
\subsubsection{Alternative Optimization}
Based on Algorithm \ref{Algorithm_1} and Algorithm \ref{Algorithm_2}, the proposed  AO algorithm that solves the two subproblems iteratively is shown in Algorithm \ref{Algorithm_3}. 
Starting from a feasible point $(\mathbf{P}^{[0]}, \mathbf{\Theta}_r^{[0]}, \mathbf{\Theta}_t^{[0]}, \mathbf{c}^{[0]}, \lambda_r^{[0]}, \lambda^{[0]},  \lambda_1^{[0]}, \lambda_2^{[0]})$, at each  iteration $[z]$, we first update   $\mathbf{\Theta}_r^{[z]}$ and $\mathbf{\Theta}_t^{[z]}$ by Algorithm \ref{Algorithm_1}.
Given $\mathbf{\Theta}_r^{[z]}$ and $\mathbf{\Theta}_t^{[z]}$, we then update $\mathbf{P}^{[z]}$, common rate and time allocation  based on Algorithm \ref{Algorithm_2}. 
By iteratively solving these two subproblems, the objective function is updated until convergence.
\begin{algorithm}[!t]	
\textbf{Initialize}:
 	 $j$, $z=0$,
   $\bm{\Theta}^{[z]}_r$, $\bm{\Theta}^{[z]}_t$,
   $\mathbf{P}^{[z]}$,
   $\mathbf{c}^{[z]}$, $\lambda_r^{[z]}$, $\lambda^{[z]}$,  $\lambda_1^{[z]}$, $\lambda_2^{[z]}$\;
\Repeat{\textit{convergence}}   
 { $z\longleftarrow z +1$\\
 Given $\mathbf{P}^{[z-1]},\mathbf{c}^{[z]}, \lambda_r^{[z]},\lambda^{[z]},\lambda_1^{[z]},\lambda_2^{[z]}$, use Algorithm \ref{Algorithm_1}
   to obtain $\bm{\Theta}_t^{[z]}$ and $\bm{\Theta}_r^{[z]}$.\\
Given $\bm{\Theta}_t^{[z]}$ and $\bm{\Theta}_r^{[z]}$, use Algorithm \ref{Algorithm_2} to obtain $\mathbf{P}^{[z]}, \mathbf{c}^{[z]},\lambda_r^{[z]},\lambda^{[z]},\lambda_1^{[z]},\lambda_2^{[z]}$.
   }
\caption{Proposed  AO algorithm }
\label{Algorithm_3}		
\end{algorithm}
\subsection{Proposed Low-complexity Algorithm}
\label{LOW algorithm}
In this subsection, we propose a low-complexity algorithm  for the STAR RIS passive beamforming and transmit  beamforming optimization subproblems.
\subsubsection{STAR RIS passive beamforming optimization}
We begin by specifying the proposed low-complexity algorithm tailored for the ES protocol. Subsequently, we extend this algorithm to the MS and TS protocols.
\par 
\textit{ES}: To develop a low-complexity algorithm for designing $\bm{\Theta}_r$ and $\bm{\Theta}_t$ in the ES protocol, we consider the following problem aimed at maximizing the sum of channel gains:
\begin{equation}
\label{Low_STAR}
		\begin{split}
			\max_{\bm{\Theta}_r,\bm{\Theta}_t}\,\, &\sum_{k=1}^K \big\|\mathbf{g}^H_k+\mathbf{h}_k^H\bm{\Theta}_i\mathbf{E}\|^2,i\in\{r,t\}\\
		\text{s.t.}\,\,	&\text{(\ref{FD-ES:4}), (\ref{FD-ES:6})}.
		\end{split}
\end{equation}
Constraints (\ref{FD-ES:4}) and (\ref{FD-ES:6}) are non-convex, which make the problem (\ref{Low_STAR}) difficult to solve.
To tackle such a non-convex problem, we first relax  constraints (\ref{FD-ES:4}) and (\ref{FD-ES:6}) to $\|\mathbf{\Theta}_r \|_F^2+ \|\mathbf{\Theta}_t \|_F^2 \leq N$, where $\|\cdot \|_F$ denotes the Frobenius norm.
This approximation relaxes the feasible solutions of $\bm{\Theta}_r$ and $\bm{\Theta}_t$ to a convex set $\mathcal{S}=\{(\bm{\Theta}_r,\bm{\Theta}_r)|\|\mathbf{\Theta}_r \|_F^2+ \|\mathbf{\Theta}_t \|_F^2 \leq N\}$.
Therefore, problem (\ref{Low_STAR}) is relaxed to
\begin{subequations}
    \label{Low_STAR:1}
    \begin{align}
        	\max_{\bm{\Theta}}\,\, &f(\bm{\Theta})\triangleq\big\|\mathbf G^H+\mathbf{H}^H\bm{\Theta} \mathbf{E}_x\|_F^2\\
\text{s.t.}\,\,	&\bm{\Theta}\in\mathcal{S}, \label{low_constraint}
    \end{align}
\end{subequations}
where $\bm{\Theta}=\begin{bmatrix}
 \bm{\Theta}_r & \mathbf{0}\\
 \mathbf{0} & \bm{\Theta}_t
\end{bmatrix}$, $\mathbf G=[\mathbf{G}_r;\mathbf{G}_t], \mathbf{G}_r=\{\mathbf{g}_k|k\in\mathcal{K}_r \},\mathbf{G}_t=\{\mathbf{g}_k| k\in\mathcal{K}_t\}$, $\mathbf H=[\mathbf{H}_r;\mathbf{H}_t], \mathbf{H}_r=\{\mathbf{h}_k| k\in\mathcal{K}_r\},\mathbf{H}_t=\{\mathbf{h}_k|k\in\mathcal{K}_t\}$ and $\mathbf{E}_x=[\mathbf{E};\mathbf{E}]$.
\par
Inspired by the method proposed in \cite{fang2023low}, we employ the gradient decent method at the point $\bm{\Theta}_0=\mathbf{0}$ to derive a closed-form solution of (\ref{Low_STAR:1}). The corresponding gradient of $f(\bm{\Theta})$ at $\bm{\Theta}_0=\mathbf{0}$ is
\begin{equation}
    \label{Low:GD}
     \bigtriangledown_{\bm{\Theta}} f(\bm{\Theta}_0)=\mathbf{H}\mathbf{G}^H\mathbf{E}_x^H,
\end{equation} 
By using (\ref{Low:GD}) as the decent direction, i.e., $\mathbf{D}_0=\bigtriangledown_{\bm{\Theta}} f(\bm{\Theta}_0)$, we then take the Armijo rule to determine the step size $\alpha$ as 
\begin{equation}
\label{Armjio}
 f(\bm{\Theta}_0+\alpha\mathbf{D}_0) \geq  f(\bm{\Theta}_0)+\phi\cdot \alpha\cdot \mathrm{tr}(\bigtriangledown_{\bm{\Theta}} f(\bm{\Theta}_0)^H\mathbf{D}_0),
\end{equation}
where $\phi\in (0,0.5)$ is a constant shrinkage factor.
Substituting (\ref{Low:GD}) into (\ref{Armjio}), we  obtain 
\begin{equation}
    \label{Armjio2}    \alpha^2\|\mathbf{H}^H\mathbf{H}\mathbf{G}^H\mathbf{E}_x^H\mathbf{E}_x\|^2_F+\alpha\cdot(2-\phi)\|\mathbf{H}\mathbf{G}^H\mathbf{E}_x^H\|^2_F\geq 0
\end{equation}
To guarantee the solution is on the boundary of the convex set $\mathcal{S}$, we set $\alpha=\frac{\sqrt{N}}{\|\mathbf H\mathbf G^H\mathbf E_x^H\|_F}$. 
Then, we obtain the following  solution of problem (\ref{Low_STAR:1}) as
\begin{equation}
\label{Low_STAR:solution}
\bm\Theta=	\frac{\sqrt{N}}{\|\mathbf H\mathbf G^H\mathbf E_x^H\|_F} \mathbf H\mathbf G^H\mathbf E_x^H.
\end{equation}
\par 
However,  the solution (\ref{Low_STAR:solution}) for the relaxed problem (\ref{Low_STAR:1}) may not satisfy the original constraints (\ref{FD-ES:4}), (\ref{FD-ES:6}). Therefore, we propose to project $\bm{\Theta}$ into the  feasible region.
Following \cite{fang2023low}, 
we first define the operator $\mathrm{sym}(\cdot)$ that projects any square matrix $\mathbf X$ to a symmetric matrix, i.e.,
\begin{equation}
\label{eq:sym}
\mathrm{sym}\left(\mathbf{X}\right)=\frac{1}{2}\left(\mathbf{X}^T+\mathbf{X}\right),
\end{equation}
and the  operator $\mathrm{uni}(\cdot)$ which projects an arbitrary  matrix $\mathbf X$ to a unitary matrix as
\begin{equation}
\label{eq:uni}
    \mathrm{uni}\left(\mathbf{X}\right)=\mathbf{UV}^H,
\end{equation}
where $\mathbf{U}$ and $\mathbf{V}$ unitary matrices obtained by the singular value decomposition (SVD) of $\mathbf{X}$, i.e.,  $\mathbf{X}=\mathbf{USV}^H$.
Consider $\mathbf{X}$ with dimension $N$ and rank $R$. Then, the unitary matrices $\mathbf{U}$ and $\mathbf{V}$ can be partitioned as $\mathbf{U}=[\mathbf{U}_R,\mathbf{U}_{N-R}]$ and $\mathbf{V}=[\mathbf{V}_R,\mathbf{V}_{N-R}]$, respectively. Based on the operators in (\ref{eq:sym}) and (\ref{eq:uni}), we could then define the symmetric unitary projection as \cite{fang2023low} 
\begin{equation}
\mathrm{symuni}\left(\mathbf{X}\right)=\mathrm{uni}\left(\mathrm{sym}\left(\mathbf{X}\right)\right)=\hat{\mathbf{U}}\mathbf{V}^H,
\end{equation}
where $\hat{\mathbf{U}}=[\mathbf{U}_R,\mathbf{V}^{*}_{N-R}]$.
\par 
Utilizing the symmetric  unitary projection, we then derive a feasible STAR RIS passive beamforming solution $\bm{\Theta}$. It's noteworthy that both  $\bm{\Theta}_r$ and $\bm{\Theta}_t$ are diagonal matrices, implying that $\bm{\Theta}$ is also diagonal. The  diagonal entries of $\bm{\Theta}$ are determined by
\begin{equation}
\label{Low_STAR:last2}
    \mathrm{diag}\{\bm{\Theta}\}=\mathrm{diag}\left\{\frac{\sqrt{2}}{2}\mathrm{symuni}\left(\mathbf H\mathbf G^H\mathbf E_x^H\right)\right\}.
\end{equation}
Since $\mathrm{symuni}\left(\mathbf H\mathbf G^H\mathbf E_x^H\right)\in\mathbb{C}^{2N\times 2N}$ is a unitary matrix, we have 
$\|\mathrm{symuni}\left(\mathbf H\mathbf G^H\mathbf E_x^H\right)\|^2_F=2N$. To ensure constraint  (\ref{FD-ES:6}), we scale each entry with  $\frac{\sqrt{2}}{2}$.
\par 
\textit{MS}: For the MS mode, we first utilize  (\ref{Low_STAR:last2}) to derive $\bm{\Theta}_r$ and $\bm{\Theta}_t$. As each element can only operate in either  reflection or transmission mode, we then compare the reflection and transmission amplitude coefficients of each STAR RIS element. The element with the higher coefficient is designated as 1, while the one with the lower coefficient is designated as 0. Based on (\ref{Low_STAR:last2}), the reflection and transmission coefficients of each STAR RIS element is designed as
\begin{equation}
\label{Low_STAR:last_MS}
\begin{split}
    \beta_{r,n}&=\left\{\begin{matrix}
 1, \textrm{ if } |\bm{\Theta}_{n,n}|-|\bm{\Theta}_{N+n,N+n}|\geq 0\\
 0, \textrm{ if }|\bm{\Theta}_{n,n}|-|\bm{\Theta}_{N+n,N+n}|< 0
\end{matrix}\right. ,\forall n\in \mathcal{N}\\
 \beta_{t,n}&=1- \beta_{r,n}.
\end{split}
\end{equation}

\par 
\textit{TS}: For the TS mode, in the reflection time slots, $\bm{\Theta}_t=\mathbf{0}$,  the amplitudes of all  elements in $\bm{\Theta}_r$ are $1$. The solution of reflection matrix is given as
\begin{equation}
\label{Low_STAR:last_TS}
\bm\Theta_r=\mathrm{diag}\{\mathrm{symuni}\left(\mathbf H_r\mathbf G_r^H\mathbf E^H\right)\}.
\end{equation}
In the transmission time slots, $\bm{\Theta}_r=\mathbf{0}$, the transmission matrix is given as 
\begin{equation}
\label{Low_STAR:last_TS2}
\bm\Theta_t=\mathrm{diag}\{\mathrm{symuni}\left(\mathbf H_t\mathbf G_t^H\mathbf E^H\right)\}.
\end{equation}

\subsubsection{Joint optimization of the active beaforming, common rate and time allocation}
With fixed $\bm{\Theta}_r$ and $\bm{\Theta}_t$, we aim to simplify the design of the remaining variables. To achieve this, we reduce the optimization dimension by fixing the direction of the precoders. Subsequently, we focus on jointly optimizing the power, common rate, and time allocation variables.
\par
Let $\bar{\mathbf{g}}_k=\mathbf{g}_k/||\mathbf{g}_k ||$, $\bar{\mathbf{G}}=[\bar{\mathbf{g}}_1,\ldots,\bar{\mathbf{g}}_K]$,  $\bar{\mathbf{p}}_k=\mathbf{p}_k/||\mathbf{p}_k ||$,  and $\bar{\mathbf{P}}=[\bar{\mathbf{p}}_1,\ldots,\bar{\mathbf{p}}_K]$.
We transform the active beamformings  in the form of $\mathbf{p}_c=\sqrt{\rho_cP_{t}}\bar{\mathbf{p}}_c, \mathbf{p}_k=\sqrt{\rho_kP_{t}}\bar{\mathbf{p}}_k$, where $\rho_c$ and $\rho_k$ are the power allocation coefficients.
For  the beamforming direction  $\bar{\mathbf{p}}_k$ of the private streams, we adopt zero-forcing (ZF) beamforming, i.e., $\bar{\mathbf{P}}=\bar{\mathbf{G}}(\bar{\mathbf{G}}^H\bar{\mathbf{G}})^{-1}$ \cite{oestges2010mimo}. 
For  the beamforming direction  $\bar{\mathbf{p}}_c$ of the common stream, we utilize maximum ratio transmission (MRT) and SVD methods as $\bar{\mathbf{p}}_c=\mathbf{u}_c$, where $\mathbf{u}_c=\mathbf{U}_c(:,1), \mathbf{G}=\mathbf{U}_c\mathbf{S}_c\mathbf{V}_c$ \cite{mao2022rate}.
Based on the aforementioned beamforming direction design, the optimization problems for all six transmission modes are simplified. In the following, we illustrate this simplification by considering the FD mode as a representative example:
\begin{subequations}
\label{low_precoder}
    \begin{align}
        \,\,\,\,&\max_{\bm{\rho},\mathbf{c}            }\,\,\min_{k\in \mathcal{K}} \,\,C_k+R^{\mathrm{FE}}_{p,k} \\
       \mbox{s.t.}\quad
	& {\textstyle \sum_{m\in\mathcal{K}}}C_m\leq R^{\mathrm{FE}}_{c,k},\forall k \in \mathcal{K}, \label{low_precoder1} \\
	&\rho_c+{\textstyle \sum_{k\in\mathcal{K}}}\rho_k\leq 1,\label{low_precoder2}
    \end{align}
\end{subequations}
where $\bm{\rho}=[\rho_c, \rho_1,\ldots,\rho_K]$ are the power allocation coefficients,  $R^{\mathrm{FE}}_{p,k}=\log_2\left(1+\rho_kP_{t}|\tilde{\mathbf{g}}^H_k\bar{\mathbf{p}}_k|^2/\sigma^2\right)$, $R^{\mathrm{FE}}_{c,k}=\log_2\left(1+\rho_cP_{t}|\tilde{\mathbf{g}}^H_k\bar{\mathbf{p}}_c|^2/(\rho_kP_{t}|\tilde{\mathbf{g}}^H_k\bar{\mathbf{p}}_k|^2+\sigma^2)\right)$. 
The optimization problem can be directly solved using the SCA-based approach specified in Algorithm \ref{Algorithm_2}.
The optimization approach for other transmission modes aligns with that of the FE mode. For FT/HE/HM/HT modes, the additional time allocation variables are jointly optimized using the SCA-based approach specified in Step 2.1.
\par
The detailed process of the proposed low-complexity algorithm to solve problems (\ref{FD-ES})--(\ref{HD-TS}) is outlined in Algorithm \ref{Algorithm_4}. Contrasted with Algorithm \ref{Algorithm_3}, the primary difference lies in the closed-form STAR RIS passive beamforming design and the closed-form active beamforming direction design. Additionally, there is no alternating optimization between the two subproblems. Each subproblem is solved once before the output.
\begin{algorithm}[!t]	
\textbf{Initialize}:$j,\mathbf{P}^{[0]}, \mathbf{c}^{[0]},\lambda_r^{[0]},\lambda^{[0]},\lambda_1^{[0]},\lambda_2^{[0]}$\;   
 Update $\bm{\Theta}_t$ and $\bm{\Theta}_r$ through (\ref{Low_STAR:last2}) if $j=\textrm{FE/HE}$, (\ref{Low_STAR:last_MS}) if $j=\textrm{FM/HM}$, (\ref{Low_STAR:last_TS}) and (\ref{Low_STAR:last_TS2}) if $j=\textrm{FT/HT}$.\\
Given $\bm{\Theta}_t$ and $\bm{\Theta}_r$,   update $\mathbf{P}$, $\mathbf{c}$, $\lambda_r$, $\lambda$, $\lambda_1$, $\lambda_2$ by solving the joint power, common rate, and time allocation problem via Algorithm \ref{Algorithm_2}.
   
\caption{Proposed low-complexity algorithm }
\label{Algorithm_4}		
\end{algorithm}
\subsection{Convergence and Computational Complexity Analysis}
\subsubsection{Convergence Analysis}
In Algorithm \ref{Algorithm_1}, the SCA method guarantees that the objective  $t^j$ monotonically increases, meaning that 
$t^{j,[z-1]}\leq t^{j,[z]}$. This arises from the fact that the  solution of $\bm{\Theta}_r$ and $\bm{\Theta}_t$ at iteration $[z-1]$ is a feasible point of the STAR RIS  problem at iteration $[z]$.
Due to STAR RIS power constraint (\ref{FD-ES:6}), there is an upper bound for $t^{j,[z]}$. Hence, the convergence of Algorithm \ref{Algorithm_1} is guaranteed.
Similarly, in Algorithm \ref{Algorithm_2}, the SCA method guarantees that  
$x^{j,[z-1]}\leq x^{j,[z]}$  since the solution  at iteration $[z-1]$ is a also feasible point of the joint  optimization problem at iteration $[z]$.
Due to the power constraint (\ref{FD-ES:5}) at the BS, there is an upper bound for the objective value. Hence, the convergence of the Algorithm \ref{Algorithm_2} is guaranteed.
The convergence of Algorithm \ref{Algorithm_1} and Algorithm \ref{Algorithm_2} implies that the objective   of the AO algorithm increases monotonically, which guarantees the convergence of Algorithm \ref{Algorithm_3}.


\subsubsection{Computational complexity of Algorithm \ref{Algorithm_3}}
\label{Complexity of the AO Algorithm}
The computational complexity of  Algorithm \ref{Algorithm_1} is $\mathcal{O}\left(N^{3.5}\log_2(1/\epsilon_1)\right)$,  where $\epsilon_1$ is the convergence tolerance.
The computational complexity of  Algorithm \ref{Algorithm_2} is $\mathcal{O}\left((KN_t)^{3.5}\log_2(1/\epsilon_2)\right)$, where $\epsilon_2$ is the corresponding convergence tolerance.
Therefore, the computational complexity of Algorithm \ref{Algorithm_3} is 
$\mathcal{O}\left(T\left(N^{3.5}\log_2(1/\epsilon_1)+(KN_t)^{3.5}\log_2(1/\epsilon_2)\right)\right)$, where $T$ denotes the number of iterations of the AO algorithm.

\subsubsection{Computational complexity of Algorithm \ref{Algorithm_4}}
For the proposed low-complexity algorithm, the computational complexity of the STAR RIS passive beamforming design is mainly occupied by the unitary projection with SVD method, which is $\mathcal{O}\left(N^3\right)$. 
For the active beamforming optimization subproblem, the computational complexity is $\mathcal{O}\left( K^{3.5}\log_2(1/\epsilon_2)\right)$.
Therefore, the overall computational complexity of Algorithm \ref{Algorithm_4} is $\mathcal{O}\left(N^3+K^{3.5}\log_2(1/\epsilon_2)\right)$, which is significantly  lower than that of Algorithm \ref{Algorithm_3}.

\begin{figure*}[tb] 
\begin{minipage}[t]{0.32\linewidth} 
\centering
\includegraphics[height=1.8in]{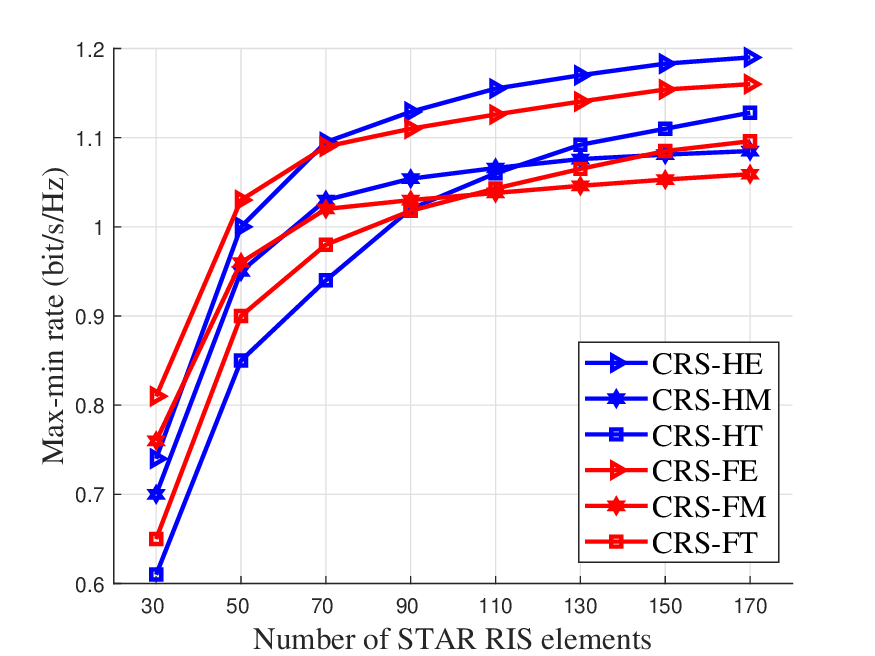} 
\caption{Max-min rate versus the number of STAR RIS elements $N$, $\mathrm{SNR}=20~\mathrm{dB}$, $K=4$, $N_t=4$.} 
\label{fig_STAR protocol} 
\end{minipage}%
\hspace{2mm}
\begin{minipage}[t]{0.32\linewidth}
\centering
\includegraphics[height=1.8in]{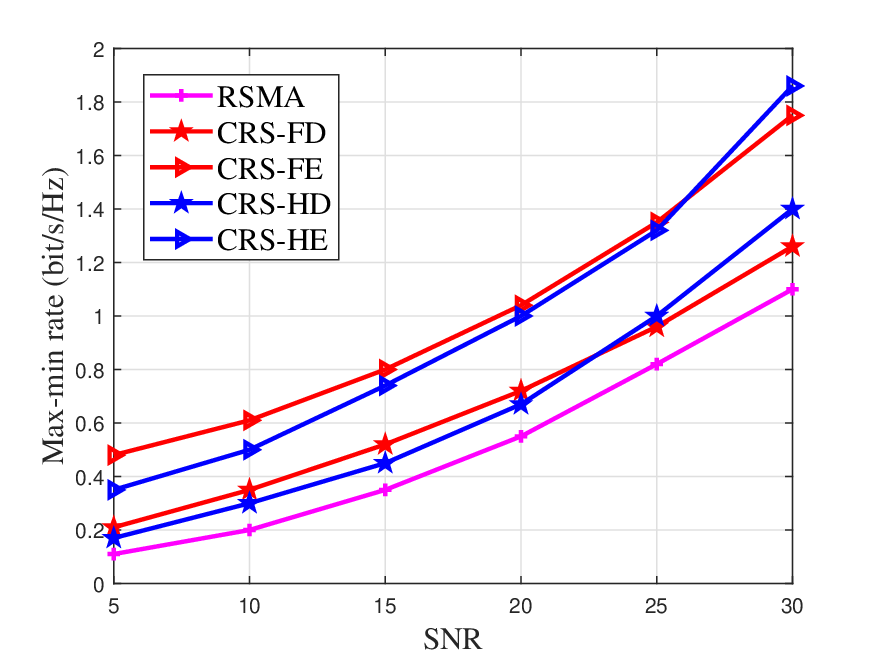}
\caption{Max-min rate versus SNR, $N=50$, \\$K=4$, $N_t=4$.}
\label{fig_Relaying}
\end{minipage}%
\hspace{2mm}
\begin{minipage}[t]{0.32\linewidth}
\centering
\includegraphics[height=1.8in]{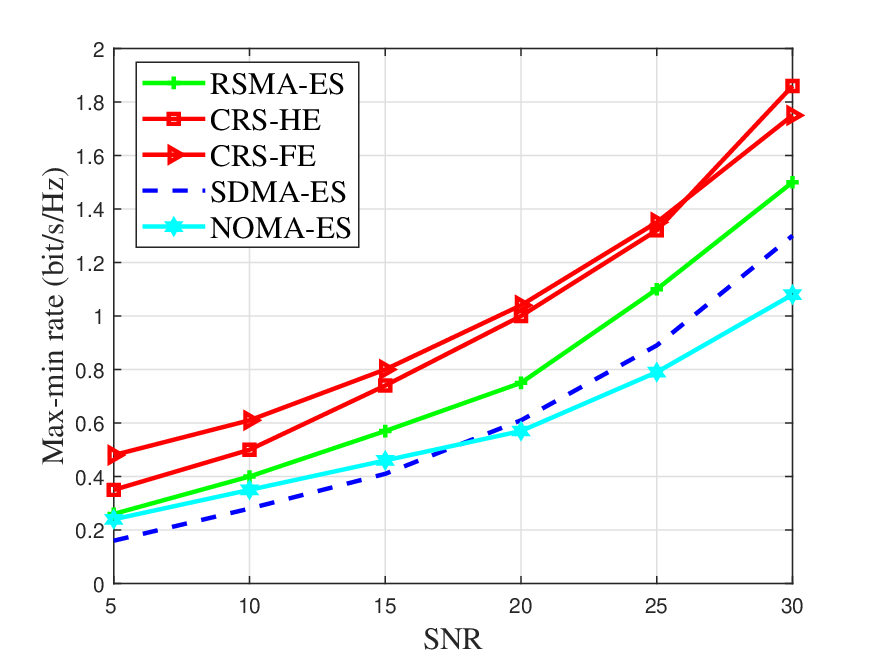}
\caption{Max-min rate versus SNR, $N=50$, \\$K=4$, $N_t=4$.}
\label{fig_MA1}
\end{minipage}
\end{figure*}

\section{Numerical Results}
\label{numerical results}

In this section, we evaluate the performance of our proposed STAR RIS-assisted CRS system framework  along with the two proposed algorithms.
\subsection{Simulation Setting}
The setting of simulation follows \cite{mu2021simultaneously}. Consider a three-dimensional (3D) space, the BS is located at $(0,0,0)$m, the STAR RIS locates at $(0,50,0)$m.
Users are randomly distributed within a circle, with the center being the location of the STAR RIS and a radius of $r=5$ meters.
We follow the method in \cite{Mao2020} to separate the users into group $\mathcal{K}_1$ and $\mathcal{K}_2$. 
We only choose one user with the largest  channel gain as the relaying user, other users are destination users. 
The channels  between BS and users, as well as between user-$m$ and user-$n$ are assumed to be Rayleigh fading, and are denoted as  $\mathbf g_k=L_{\mathrm{BU},k}\bar{\mathbf{g}}_k, h_{m,n}=L_{\mathrm{UU}}\bar{h}_{m,n}$, where $L_{\mathrm{BU},k}$ and $L_{\mathrm{UU}}$ denote the path loss between the BS and users as well as the path loss between users.  
$\bar{\mathbf{g}}_k$ and $\bar h_{m,n}$ follow the complex Gaussian distribution with a certain variance, i.e., $\bar{\mathbf{g}}_k  \sim\mathcal{CN}(0,\sigma^2_{k}),\bar h_{m,n}\sim\mathcal{CN}(0,\sigma^2_{m,n})$.
The channels between the BS and STAR RIS, as well as between STAR RIS and users are assumed to  
follow the Rician fading model with the Rician factor being 3dB.
The path loss for all the aforementioned channels is defined as
$L_x=L_0(d_x/d_0)^{-\alpha_x},x\in\{\mathrm{BU},\mathrm{UU},\mathrm{BR},\mathrm{RU}\}$, where $d_0=1$m is the reference distance, $L_0=-30$ dB is the path loss at reference distance, $d_x$ is the distance of different channels, and $\alpha_x$ denotes different channel path exponents. 
The channel path loss exponents between the BS and STAR RIS, as well as between STAR RIS and users  are 2.2 \cite{zuo2022joint}. 
The channel path loss exponent between the BS and users is 3.76 \cite{yang2021energy}.
Besides, the relaying power is $P_k=0.5P_t$, the convergence tolerance is $\epsilon=10^{-3}$, and the noise power is $\sigma=-90$ dBm.
All results are obtained by averaging over 100  random channel realizations. 
The variances of user channels are set to $\sigma^2_{k}=1$ for the relaying user, $\sigma^2_{k}=0.3$ for the destination users, and $\sigma^2_{m,n}=1$.

\par
The following  schemes are compared in this section:
\begin{itemize}
    \item \textbf{CRS-FE/FM/FT/HE/HM/HT}: This refers to the STAR RIS-assisted CRS for six different transmission modes, as we proposed in Section  \ref{system}.    
    \item \textbf{CRS-HD/FD}: This refers to the existing CRS transmission schemes based on the HD or FD protocols without the assistance of STAR RIS \cite{Mao2020, li2021full}.
    \item \textbf{RSMA-ES}: This refers to the existing STAR RIS assisted RSMA transmission scheme without user relaying \cite{katwe2023improved}. Here we only consider the ES protocol  since  it is more general than MS and TS protocols.
    \item \textbf{SDMA-ES}: This refers to the existing STAR RIS assisted SDMA transmission scheme. Again, we only consider the ES protocol.
    \item \textbf{NOMA-ES}: This refers to the existing STAR RIS assisted NOMA transmission scheme \cite{mu2021simultaneously} with the ES protocol.
     \item \textbf{RSMA}: This refers to the conventional 1-layer RSMA transmission scheme, which does not involve user relaying or assistance from STAR RIS \cite{mao2022rate}.
\end{itemize}

\subsection{Simulation Results}
\subsubsection{Comparison among STAR RIS operating 
 protocols}
From Fig. \ref{fig_STAR protocol},  we observe that 
the FE/HE mode consistently   achieves a higher max-min rate than the FM/HM and FT/HT mode. This is because the ES protocol is more general than the MS and TS protocol, and it ensures that each STAR RIS element has a larger tunable range of the amplitude coefficient.
As the number of STAR RIS elements increases, the max-min rate performance gain between FM/HM and FT/HT modes decreases. Surprisingly, when $N$ is larger than $110$, the FT/HT mode achieves a higher max-min rate than the FM/HM mode. This indicates that the TS protocol outperforms  the MS protocol when the STAR RIS element is large. With an increasing number of STAR RIS elements, the performance improvement gained from having all elements operating within the same transmission/reflection space  is outweighed by the performance degradation caused by partial time service.
\par Due to the superior max-min rate performance of ES compared to MS and TS, in the following simulation results, we only illustrate the proposed STAR RIS-assisted CRS in FE and HE modes for clarity.
\subsubsection{Comparison between  FD and HD relaying protocols}
 Fig. \ref{fig_Relaying} illustrates the max-min rate performance versus transmit SNR. We compare our proposed  CRS-FE/HE with the conventional RSMA scheme and the conventional CRS schemes with HD and FD protocols. Algorithm \ref{Algorithm_3} is employed to optimize the corresponding problems of the proposed schemes. We set $N=50, K=4,N_t=4$. 
The max-min rate increases with SNR for all transmission schemes. It is noteworthy that when SNR is below $25$ dB, the CRS-FE scheme achieves a higher max-min rate than the CRS-HE scheme. Conversely, when the SNR exceeds $25$ dB, the CRS-HE scheme outperforms the CRS-FE scheme. This divergence stems from the introduction of self-interference in FD relaying. As the SNR increases, the level of self-interference also rises. This indicates that the HD protocol is preferred in the high SNR regime while the FD protocol is preferred in the low and moderate SNR regimes.
For consistently ensuring the optimal max-min rate performance gain, we can employ FD relaying at low and moderate SNR and HD relaying at high SNR.
\par
From  Fig. \ref{fig_Relaying}, it is obvious that with the assistance of STAR RIS and CRS, CRS-FE/HE achieves explicit max-min rate gain over conventional schemes. Specifically, CRS-FE demonstrates an average relative performance gain of 117.1\% and 52.4\% over conventional RSMA and CRS-FD, respectively. This demonstrates the significant performance enhancement achieved by integrating STAR RIS and CRS.
 

\subsubsection{Comparison among different MA schemes}

In Fig. \ref{fig_MA1}, we compare various STAR RIS-assisted transmission schemes with the ES protocol when $N=50$, $K=4$, $N_t=4$.
The results show that our proposed STAR RIS-aided CRS scheme outperforms other MA schemes. It  achieves an average max-min rate gain of 42.6\%  over the RSMA-ES scheme, 90.9\% over the SDMA-ES scheme, and  77.2\% over the NOMA-ES scheme.
Besides, the RSMA-ES scheme outperforms NOMA-ES and SDMA-ES. By splitting the user message into common and private parts, RSMA facilitates more flexible interference management in all SNR regimes.

\begin{figure}[!t]
	\centering
	\subfigure[Max-min rate versus $N$]{
		\includegraphics[width = 0.475\linewidth]{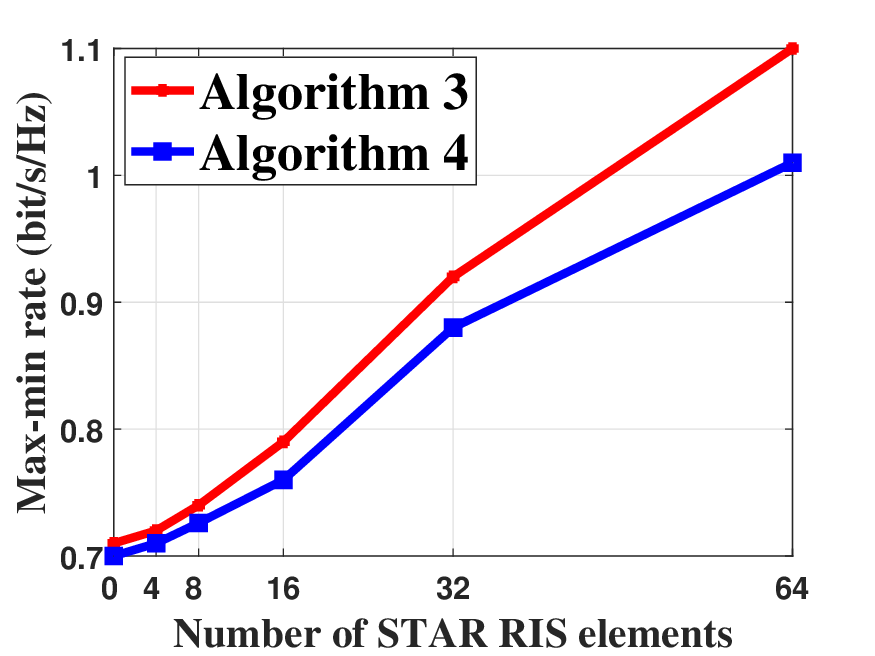}
		\vspace{-2em}
	}
	\hspace{-1em}
	\subfigure[Average CPU time versus $N$]{
		\includegraphics[width = 0.475\linewidth]{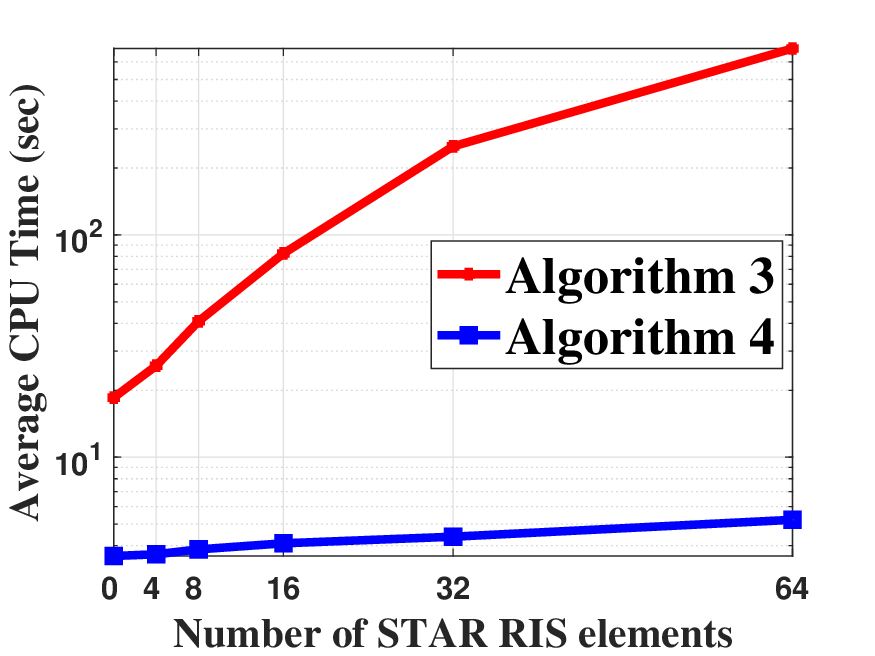}
		\vspace{-2em}
	}
	\caption{The performance of the two proposed algorithms.}
	\label{fig_Low complexity}
    \vspace{-1em}
\end{figure}

\subsubsection{Comparison between the two proposed algorithms}
In Fig. \ref{fig_Low complexity}, we compare the two proposed optimization algorithms for the proposed STAR RIS-aided CRS in the ES mode.  We set $\mathrm{SNR}=20$ dB, $K=4$, $N_t=4$.
In comparison to the proposed AO algorithm, the low-complexity algorithm reduces the average CPU time by 99.2\% while incurring only a 9.1\% performance loss at $N=64$. This confirms the effectiveness  of the proposed low-complexity algorithm.
This efficiency is primarily due to the low-complexity algorithm eliminating the need for CVX operations to update $\bm{\Theta}_r$ and $\bm{\Theta}_t$, along with the alternative optimization between two subproblems. Instead, CVX is exclusively  used for joint power,  common rate, and time allocation, leading to a significant reduction in optimization dimension.
The results suggest that if the system can accommodate high time complexity, Algorithm \ref{Algorithm_3} should be employed to achieve near-optimal max-min rate performance. Conversely, if the system is time sensitive, the low-complexity Algorithm is recommended.
\section{Conclusion}
\label{conclusion}
In this paper, we propose a novel STAR RIS-assisted  CRS transmission with six different transmission modes, namely, HE, HM, HT, FE, FM, FT.  With the objective of maximizing the minimum user rate, we then propose a unified SCA-based AO algorithm  to optimize the BS active beamforming and the STAR RIS passive beamforming iteratively under the transmit power constraint at the BS and the law of energy conservation at the STAR RIS. Meanwhile, we propose a novel low-complexity resource allocation algorithm that designs the BS active beamforming and the STAR RIS passive beamforming in closed form. Numerical results show that our proposed STAR RIS aid-CRS system achieves superior max-min rate performance gain over the existing CRS schemes and the STAR RIS-aided  MA schemes.
Furthermore, we show that the FD relaying has better max-min rate performance at low and moderate SNR while the HD relaying  is preferred at high SNR.
The ES protocol offers superior max-min rate performance compared to the MS and TS protocols, albeit with increased hardware complexity.
With a large number of STAR RIS elements, the TS protocol demonstrates a significantly higher max-min rate gain compared to the MS protocol.
Our proposed AO algorithm is better suited for systems that can accommodate higher time complexity, while the low-complexity algorithm is preferable for systems that are more time sensitive.
\bibliographystyle{IEEEtran}  
\bibliography{reference}

\end{document}